\documentclass[aps,twocolumn,prl,nofootinbib,superscriptaddress]{revtex4-1}

\received{6 Jul. 2018}
\revised{11 Oct. 2018}
\accepted{2 Nov, 2018}

\usepackage{stackengine}
\usepackage{scrextend}
\usepackage{epsfig,amsmath,amssymb,bm,color,graphicx}
\usepackage{rotating,booktabs}
\usepackage{xcolor}
\usepackage{colortbl}
\usepackage{lineno}
\usepackage{marginnote}
\usepackage{soul}
\begin{document}
\newcommand{\Fermi}{\textit{Fermi} }
\newcommand{\FermiLAT}{\textit{Fermi}-LAT }

\title{Unresolved Gamma-Ray Sky through its Angular Power Spectrum}

\author{M.~Ackermann}
\affiliation{Deutsches Elektronen Synchrotron DESY, D-15738 Zeuthen, Germany}
\author{M.~Ajello}
\affiliation{Department of Physics and Astronomy, Clemson University, Kinard Lab of Physics, Clemson, SC 29634-0978, USA}
\author{L.~Baldini}
\affiliation{Universit\`a di Pisa and Istituto Nazionale di Fisica Nucleare, Sezione di Pisa I-56127 Pisa, Italy}
\author{J.~Ballet}
\affiliation{Laboratoire AIM, CEA-IRFU/CNRS/Universit\'e Paris Diderot, Service d'Astrophysique, CEA Saclay, F-91191 Gif sur Yvette, France}
\author{G.~Barbiellini}
\affiliation{Istituto Nazionale di Fisica Nucleare, Sezione di Trieste, I-34127 Trieste, Italy}
\affiliation{Dipartimento di Fisica, Universit\`a di Trieste, I-34127 Trieste, Italy}
\author{D.~Bastieri}
\affiliation{Istituto Nazionale di Fisica Nucleare, Sezione di Padova, I-35131 Padova, Italy}
\affiliation{Dipartimento di Fisica e Astronomia ``G. Galilei'', Universit\`a di Padova, I-35131 Padova, Italy}
\author{R.~Bellazzini}
\affiliation{Istituto Nazionale di Fisica Nucleare, Sezione di Pisa, I-56127 Pisa, Italy}
\author{E.~Bissaldi}
\affiliation{Dipartimento di Fisica ``M. Merlin" dell'Universit\`a e del Politecnico di Bari, I-70126 Bari, Italy}
\affiliation{Istituto Nazionale di Fisica Nucleare, Sezione di Bari, I-70126 Bari, Italy}
\author{R.~D.~Blandford}
\affiliation{W. W. Hansen Experimental Physics Laboratory, Kavli Institute for Particle Astrophysics and Cosmology, Department of Physics and SLAC National Accelerator Laboratory, Stanford University, Stanford, CA 94305, USA}
\author{R.~Bonino}
\affiliation{Istituto Nazionale di Fisica Nucleare, Sezione di Torino, I-10125 Torino, Italy}
\affiliation{Dipartimento di Fisica, Universit\`a degli Studi di Torino, I-10125 Torino, Italy}
% ...
\collaboration{The \textit{Fermi}-LAT Collaboration}
\author{S.~Ammazzalorso}
\affiliation{Istituto Nazionale di Fisica Nucleare, Sezione di Torino, I-10125 Torino, Italy}
\affiliation{Dipartimento di Fisica, Universit\`a degli Studi di Torino, I-10125 Torino, Italy}
\author{N.~Fornengo}
\affiliation{Istituto Nazionale di Fisica Nucleare, Sezione di Torino, I-10125 Torino, Italy}
\affiliation{Dipartimento di Fisica, Universit\`a degli Studi di Torino, I-10125 Torino, Italy}
\author{M.~Regis}
\affiliation{Istituto Nazionale di Fisica Nucleare, Sezione di Torino, I-10125 Torino, Italy}
\affiliation{Dipartimento di Fisica, Universit\`a degli Studi di Torino, I-10125 Torino, Italy}

\begin{abstract}
The gamma-ray sky has been observed with unprecedented accuracy in the last decade by the \Fermi large area telescope (LAT), allowing us to resolve and understand the high-energy Universe.
The nature of the remaining unresolved emission (unresolved gamma-ray background, UGRB) below the LAT source detection threshold can be uncovered by characterizing the amplitude and angular scale of the UGRB fluctuation field. This work presents a measurement of the UGRB autocorrelation angular power spectrum  based on eight years of \Fermi LAT Pass 8 data products. The analysis is designed to be robust against contamination from resolved sources and noise systematics. The sensitivity to subthreshold sources is greatly enhanced with respect to previous measurements.
We find evidence (with $\sim3.7\sigma$ significance) that the scenario in which two classes of sources contribute to the UGRB signal is favored over a single class. A double power law with exponential cutoff can explain the anisotropy energy spectrum well, with photon indices of the two populations being $2.55\pm0.23$ and $1.86\pm0.15$.
\end{abstract}

\maketitle

\section{I. Introduction} \label{sec:intro}
The Universe has a network of structures. The so-called cosmic web was formed by gravitational instabilities, starting from the tiny density fluctuations that originated during primordial inflation, which evolved into structures at very different scales, from stars to galaxies, up to galaxy clusters and filaments. Furthermore, this texture nurtures the formation of non-thermal astronomical sources.\\
In ten years of operation, the {\it Fermi} Large Area Telescope (LAT) has been providing an unprecedented census of non-thermal emitters in gamma rays. The most recent {\it Fermi}-LAT 8-year preliminary Point Source List (FL8Y\footnote{\url{https://fermi.gsfc.nasa.gov/ssc/data/access/lat/fl8y/}}) contains 5524 objects detected with a significance greater than $4\sigma$ between $100$ MeV and $1$ TeV.\\
Gamma-ray sources that are too dim to be resolved individually by {\it Fermi}-LAT contribute cumulatively to the Unresolved Gamma-Ray Background (UGRB), see Ref. \citep{Fornasa:2015qua} for a recent review.
Although the exact composition of the UGRB is still an open issue, high-latitude sources are expected to be mostly of extragalactic origin. Therefore they should follow the matter potential in the Universe (with some bias) and should be distributed anisotropically in the sky.\\
Different populations of gamma-ray emitters induce anisotropies in the UGRB with different amplitudes and different angular and energy spectra \citep{2009PhRvL.102x1301S}. A measurement of the gamma-ray angular power spectrum (APS) can therefore constrain the nature of the UGRB in a complementary way with respect to the intensity energy spectrum and the 1-point photon count probability distribution \citep{2011ApJ...738..181M, 2015JCAP...09..027F, 2016ApJS..225...18Z, 2016ApJ...832..117L, 2016ApJ...826L..31Z, DiMauro:2017ing}. A different but related approach based on two-point statistics is the cross correlation of the gamma-ray sky with independent probes tracing the large scale structures of the Universe \citep{xia11,2014PhRvD..90b3514A,Xia:2014,Regis:2015zka,2015ApJS..221...29C,2014JCAP...10..061A,2015PhRvD..92l3540S,2016JCAP...06..045A, 2017ApJS..228....8B,2017arXiv170809385L,2017arXiv170900416L,Camera:2012cj,Camera:2014rja,Shirasaki:2014noa,2016PhRvD..94f3522S,2017MNRAS.467.2706T,Fornengo:2014cya}.\\
The first detection of anisotropies in the UGRB was reported by the {\it Fermi}-LAT Collaboration in 2012 \citep{Ackermann:2012uf}, and then updated in 2016, employing 81 months of Pass 7 {\it Fermi} LAT data from 0.5 to 500 GeV \citep{2016PhRvD..94l3005F} (hereafter Fornasa et al.). The latter analysis revealed a hint that the measured APS might be due to more than one population of sources \citep{2017PhRvD..95l3006A}.\\
The raw APS (namely, the one that is measured directly from {\it Fermi}-LAT gamma-ray maps) is the sum of three contributions: a) a noise term, $C_N$, due to fluctuations of photon counts, showing no correlation between different pixels in the sky and thus producing a flat APS; b) the auto-correlation of fluctuations due to individual sources with themselves ($C_P$): in the limit of point-like sources and infinite angular resolution of the telescope, this term shows up only at zero angular separation in real space (which implies a flat APS), but the finite size of the point-spread function (PSF) makes the associated APS decrease at high multipoles; c) the correlation between fluctuations induced by sources located in different positions in the sky: this contribution is expected to trace the cosmic web.\\
$C_N$ is expected to become less and less relevant as the statistics grow. $C_P$ decreases as the brightest sources become resolved. In the current state of gamma-ray searches, it is still the dominant physical contribution to the APS. The third term is expected to eventually take over once the sensitivity of the telescope is such that a sufficiently large number of bright sources are resolved (and so no longer contribute to the UGRB).\\

\section{II. Signal extraction} \label{sec:Extract}
\begin{figure}[t]
\centering
\includegraphics[width=8cm]{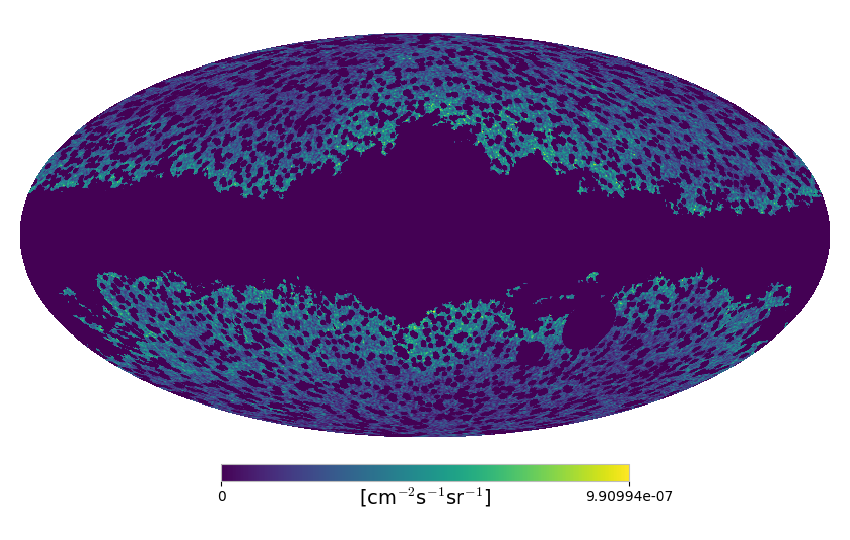}
\includegraphics[width=8cm]{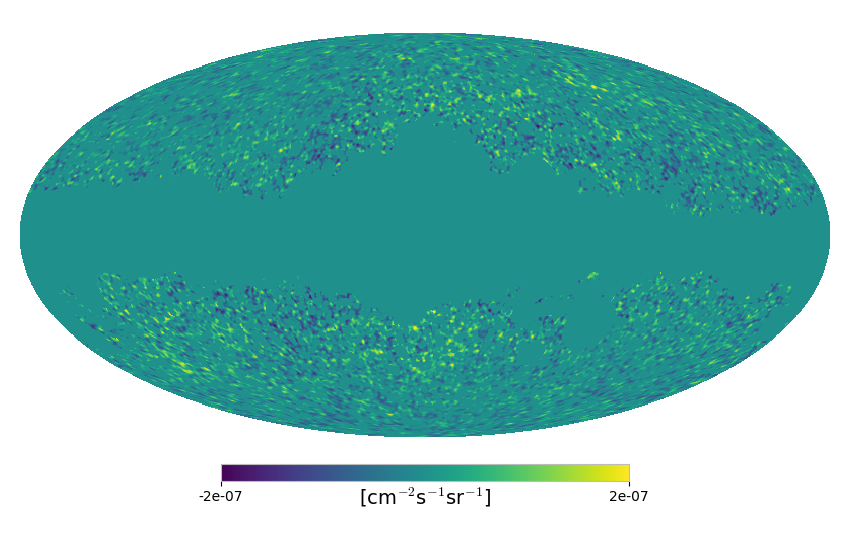}
\caption{Top: Mollweide projection of the all-sky intensity map for photon energies in the (1.7--2.8) GeV interval, after the application of the mask built for this specific energy bin. Bottom: Mollweide projection of the UGRB map between (1.7--2.8) GeV. Masked pixels are set to 0; Maps have been downgraded to order 7 for display purposes and smoothed with a Gaussian beam with $\sigma = 0.5^{\circ}$ and $\sigma = 1^{\circ}$ respectively.}
\label{fig:maps}
\end{figure}

\bigskip
A study of morphological anisotropies requires data with a good angular resolution. The data selection used in this analysis is designed to obtain the purest event sample and to maximize both the precision of the reconstructed arrival directions and the total photon counts statistics. For these reasons we select Pass 8 \footnote{\url{https://fermi.gsfc.nasa.gov/ssc/data/analysis/documentation/Cicerone/Cicerone_Data/LAT_DP.html}} data of the P8R3\_SOURCEVETO\_V2 event class\footnote{The new SOURCEVETO event class, currently under development in the LAT collaboration and planned for public release, has an acceptance comparable to P8R2\_CLEAN\_V6 with a residual contamination almost equal to that of P8R2\_ULTRACLEANVETO\_V6 at all energies.}, and we reject the quartile of events with the worst PSF, which corresponds to all the events flagged as PSF0 type.\\
The data selection comprises 8 years and is performed using version v10r0p5 of the \textit{Fermi} Science Tools. Data in the energy range between 100 MeV and 1 TeV is subdivided into 100 logarithmically spaced ``micro" bins, and for each of them we produce a count map and an exposure map, whose ratio gives 100 flux maps.
They are then summed in order to obtain intensity maps in 12 ``macro" energy bins between 524 MeV and 1 TeV (see Tab.~\ref{tab:tab1}). This choice minimizes the effects of the energy dependence of the exposure, and we exploited this fine binning in the estimation of the autocorrelation as will be explained in the next section. Data are spatially binned with HEALPix \citep{HealpixRef} order 9.\\
The flux maps are masked such that the majority of the Galactic interstellar emission is removed, as well as the contribution from the resolved sources listed in the FL8Y source list (adding sources from the 3FHL catalog \citep{TheFermi-LAT:2017pvy} when considering energies beyond 10 GeV). The source mask is built taking into account both the brightness of each source and the energy dependence of the PSF. We tested the effectiveness of our masks performing several tests described in the Supplemental Online Material (SOM)~\citep{refSOM}.
Fig.~\ref{fig:maps} illustrates the mask built for the energy bin between 1.7 and 2.8 GeV.\\
In order to eliminate the residual Galactic contribution, we subtract the Galactic diffuse emission (GDE) with the model gll\_iem\_v6.fits described in \citep{Acero:2016qlg}: in each micro energy bin, we perform a Poissonian maximum likelihood fit of data maps (considering only unmasked pixels) with the GDE model (with a free normalization) and a spatially constant term accounting for the UGRB and possible cosmic-ray residuals in the LAT; we find normalizations compatible with one within $1\sigma$ statistical uncertainty in each energy bin, and then we subtract the normalized GDE model from data maps. An example of masked map leaving only the UGRB in the energy bin (1.7--2.8) GeV is illustrated in the right panel of Fig.~\ref{fig:maps}.

\onecolumngrid

\twocolumngrid
\section{III. Angular power spectrum analysis} \label{sec:APS}
The APS of intensity fluctuation is defined as: $C_{\ell}^{ij} = \frac{1}{2\ell+1}\langle \sum_m a_{\ell m}^{i} a_{\ell m}^{j*}\rangle$, where the brackets indicate the average on the modes $m$, the indexes $i$ and $j$ label the $i^{\rm{th}}$ and the $j^{\rm{th}}$ energy bins. When $i=j$, we refer to autocorrelation, to cross-correlation otherwise. The coefficients $a_{\ell m}$ are given by the expansion in spherical harmonics of the intensity fluctuations, $\delta I_g(\vec n) =  \sum_{\ell m} a_{\ell m} Y_{\ell m}(\vec n)$, with $\delta I_g(\vec n)\equiv I_g (\vec n) - \langle I_g \rangle$ and $\vec n$ identifies the direction in the sky.The APS hence quantifies the amplitude of the anisotropy associated with each multipole $\ell$, which roughly corresponds to a pattern ``spot" size of $\lambda \simeq (180^{\circ}/{\ell})$.\\
We compute the APS with PolSpice \citep{szapudi01,chon04}, a Fortran90 software tool which is based on the fast Spherical Harmonic Transform. PolSpice estimates the covariance matrix of the different multipoles taking into account the correlation effect induced by the mask with the algorithm described in \citep{efstathiou04,challinor05}.
Prior to the measurement, we exploited the standard HEALPix routine to removed the monopole and the dipole terms from the intensity maps in order to eliminate possible spectral leakage (owing to the masking) of these large-scale fluctuations (which have large amplitudes) on the small scales we are interested in.\\
The resolution of the maps and the effect of the PSF are accounted for respectively by the pixel window function, $W^{{\rm pix}}(\ell)$, and the beam window function, $W^{{\rm beam}}(E,\ell)$, whose computation is described in the SOM.
Any random noise would contribute to the signal when the autocorrelation  in the $i^{\rm{th}}$ energy bin, $C_\ell\equiv C_\ell^{ii}$, is performed, hence it must be subtracted from the raw APS. We know that a Poissonian white noise would have a flat APS which can be estimated as in Fornasa et al.:  $C_N = \frac{\langle n^i_{\gamma,{\rm pix}}/(\epsilon^i_{{\rm pix}})^2 \rangle} {\Omega_{{\rm pix}}}$, $n^i_{\gamma,{\rm pix}}$ being the photon counts in the $i$th pixel, $\epsilon^i_{{\rm pix}}$ the exposure, $\Omega_{{\rm pix}}$ the pixel solid angle, and the average is on the unmasked pixels. Considering this as the only noise term, any other random component not following a Poisson distribution would not be taken into account. Moreover, the above equation for $\hat C_N$ represents only an estimator of the true $C_N$. 
 Indeed, we found evidence of an underestimation of the noise term above a few GeV, and devised a method to determine the autocorrelation APS without relying on the estimate of $C_N$. We exploit cross-correlations between different but closely adjacent micro energy bins: these are not affected by the noise term, since any kind of noise would not correlate between independent data samples. Also, we do not expect any effect due to the energy resolution of the instrument since the width of the micro bins is larger than the energy resolution, except for bins below 1 GeV (the first macro bin) whose result is anyway compatible with the one obtained by the standard autocorrelation method which is valid at those energies.
As explained in the previous section, our macro energy bins are composed of a number $N_b$ of micro energy bins. The APS computed in the macro bin can be seen as the sum of all the auto and cross APS computed for all the micro energy bins:
\begin{equation}
C_{\ell} = \sum_{\alpha=1}^{N_b} C^{\alpha\alpha}_{\ell,\rm{micro}} + 2 \sum_{\substack{\alpha,\beta \\ \alpha>\beta}} C^{\alpha\beta}_{\ell,\rm{micro}}
\label{eq:exact}
\end{equation}
where $\alpha, \beta = 1, ..., N_b$.\\
Under the reasonable assumption that the contributing sources have a broad and smooth energy spectrum, the APS for each macro energy bin can be obtained as:
\begin{equation}
C_{\ell}= \frac{N_b}{N_b -1} \sum_{\substack{\alpha,\beta \\ \alpha\neq\beta}}\frac{C^{\alpha\beta,\rm{Pol}}_{\ell,\rm{micro}}}{W_{E_{\alpha}}(\ell)W_{E_{\beta}}(\ell)}
\label{eq:improved}
\end{equation}
where $W_{E_{\alpha}}(\ell) = W^{{\rm beam}}_{E_{\alpha}}(\ell)W^{{\rm pix}}_{E_{\alpha}}(\ell)$ and $N_b$ is the number of micro bins in each macro energy bin\footnote{Note that Eq.~\ref{eq:improved} returns a better approximation if the width of the micro bins decreases, and/or $N_b$ increases, and/or the global spectrum of the underlying source population flattens. We calculated that when $N_b>3$, considering our micro energy bin width and an anisotropy energy spectrum $\sim E^{-4}$, the difference between Eq.~\ref{eq:exact} and Eq.~\ref{eq:improved} is less than 1\%. 
We use $N_b$ = 6 for all but the two highest-energy macro bins, for which we use $N_b$ = 11 and $N_b$ = 12, respectively.}. In this way, we avoid relying on the autocorrelation of the micro bins and therefore on the estimate of the noise. The SOM provides more details to support this approach.
\begin{figure}[h]
\centering
\includegraphics[width=9.5cm]{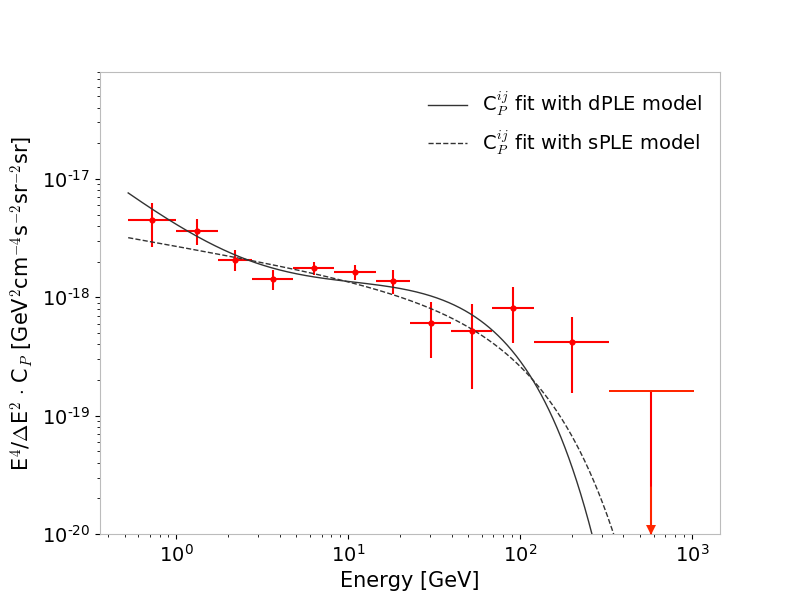}
\caption{Anisotropy energy spectrum C$_P$(E), whose values are reported in Tab.~\ref{tab:tab1}. We also show the best-fit models sPLE (single power law with exponential cutoff) and dPLE (double power law with exponential cutoff), and we stress that they have been obtained by considering the total set of C$_P^{ij}$ from both auto- and cross-correlations between macro energy bins (see the last section for details about the fitting procedure).}
\label{fig:Cp}
\end{figure}

\bigskip
\section{A. Autocorrelation \\anisotropy energy spectrum} \label{subsec:Auto}
For each energy bin, we find no evidence for an $\ell$-dependent APS. This flat behavior is expected if the anisotropy signal is dominated by unresolved point-like sources isotropically distributed in the sky. 
We therefore derive the level of anisotropy, $C_P$, for each energy bin by fitting the APS with a constant value: this provides the energy spectrum of the anisotropy signal due to gamma-ray point-like sources. Prior to this fit, each APS was binned to reduce the correlation among neighboring $C_{\ell}$. To carry out the binning in the most effective way, we implemented the unweighted averaging procedure proposed in Fornasa et al., which was validated with Monte Carlo simulations (see Sec. IV-A of Fornasa et al.). The range of multipoles considered for the fitting procedure is determined taking into account several considerations: we exclude $l < 50$ where residual large-scale contributions from the foreground emission are significant and leakage from large-scale fluctuations still could be important; the beam window function correction is inaccurate when considering scales much smaller than the PSF: the upper limit in multipole depends on the PSF and on the photon statistics at a specific energy, and hence varies with the energy bin. Further details are provided in the SOM.\\
In Tab.~\ref{tab:tab1}, we report the obtained $C_P$ as a function of energy, as well as the fitting range of multipoles considered, and the systematics related to the uncertainty of the \FermiLAT effective area $A_{\rm{eff}}$\footnote{This uncertainty is obtained doubling the systematic uncertainty of the instrumental $A_{\rm{eff}}$, since the APS is the square of the intensity. \url{https://fermi.gsfc.nasa.gov/ssc/data/analysis/LAT\_caveats.html}}.\\
Fig.~\ref{fig:Cp} shows our measurement of the anisotropy energy spectrum between 524 MeV and 1 TeV.

\begin{table}[t]
\centering
\begin{tabular}{ c|c|c|c }
\hline
\hline
E$_{min}$-E$_{max}$&Fit range&C$_P\pm\delta$C$_P$&C$_{P,A_{\rm{eff}}}^{sys}$\\
$[{\rm GeV}]$&[l$_{min}$-l$_{max}$]&[${\rm cm}^{-4}{\rm s}^{-2}{\rm sr}^{-2}{\rm sr}$]&[$\%$]\\
\hline
$0.5-1.0$&$50-150$&$(3.7\pm 1.5)$ E-18&20\\
$1.0-1.7$&$50-250$&$(6.6\pm 1.6)$ E-19&20\\
$1.7-2.8$&$50-450$&$(9.4\pm 1.8)$ E-20&20\\
$2.8-4.8$&$50-600$&$(3.4\pm 0.63)$ E-20&20\\
$4.8-8.3$&$50-900$&$(1.4\pm 0.18)$ E-20&20\\
$8.3-14.5$&$50-1000$&$(4.3\pm 0.61)$ E-21&20\\
$14.5-22.9$&$50-1000$&$(9.0\pm 2.1)$ E-22&20\\
$22.9-39.8$&$50-1000$&$(2.1\pm 1.0)$ E-22&20\\
$39.8-69.2$&$50-1000$&$(5.9\pm 4.0)$ E-23&20\\
$69.2-120.2$&$50-1000$&$(3.1\pm 1.5)$ E-23&22\\
$120.2-331.1$&$50-1000$&$(1.2\pm 0.73)$ E-23&25\\
$331.1-1000.0$&$50-1000$&$(-4.4\pm 11)$ E-25&32\\
\hline
\hline
\end{tabular}
\caption{C$_P$ values and the corresponding errors $\delta$C$_P$ for each energy bin, as well as the range of multipoles considered in the fit of the APS and the systematic error associated to the instrumental effective area.}
\label{tab:tab1}
\end{table}

\bigskip
\section{B. Cross-correlations \\between energy bins}
\begin{figure}[t]
\centering
\includegraphics[width=8.5cm]{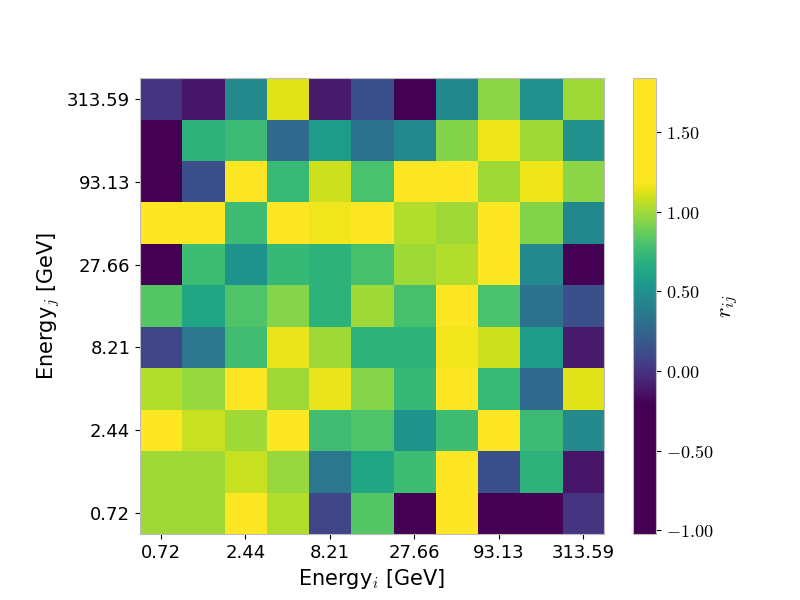}
\quad
\includegraphics[width=7.6cm]{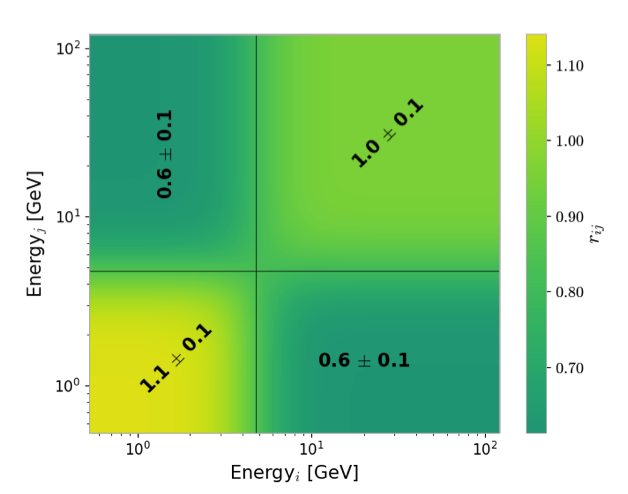}
\caption{Top: Cross-correlation coefficient $r_{ij}$ matrix. This matrix is symmetric and has 1 on the diagonal by construction; the column and the row involving the last energy bin have been removed since the autocorrelation value is negative there and the corresponding $r_{ij}$ values have negative roots. Bottom: mean values and standard deviation of the mean in each sub-rectangle of the $r_{ij}$ matrix. If only one population contributed to the anisotropy signal, the mean values in the off-diagonal sub-rectangles would be values compatible with one, which is not the case.}
\label{fig:rij}
\end{figure}

A way to discriminate whether the signal is due to either a single class or multiple classes of point-like sources is to study the cross-correlations among energy bins: distinct populations of sources, presenting different energy spectra, reasonably lie in different sky positions. \\
Similarly to the autocorrelation APS, we find flat cross-APS when performing cross-correlations between macro energy bins. If the anisotropy cross signal is due to a single class of sources, then $C_P^{ij} = \sqrt{C_P^{ii} \ C_P^{jj}}$, where $C_P^{ii}$ and $C_P^{jj}$ are the autocorrelation anisotropy levels in the energy bins $i$ and $j$ respectively. The ratio $r_{ij} = C_P^{ij}/\sqrt{C_P^{ii} \ C_P^{jj}}$  is the cross-correlation coefficient: it should be compatible with 1 for each $ij$ pair if the signal is due to a single class of sources. Fig.~\ref{fig:rij} (left panel) shows the $r_{ij}$ matrix: low-energy bins clearly correlate with nearby bins, while correlate less with the high-energy ones, and \textit{vice versa}, meaning that sources contributing to the signal at low energy are not located at the same positions (on the spherical sky projection along the line of sight) as those that contribute at high energy. Hence, more than one class of source is present.  

\bigskip
\section{VI. Discussion} \label{sec:discuss}
The global measurement, given by both the auto and the cross-correlations, can be exploited to perform a statistical test, in order to establish whether a double-population scenario is favored with respect to a single-population case. We compute the $\chi^2$ for  two models: a single power law with an exponential cutoff, sPLE (3 free parameters: normalization, spectral index and cutoff energy), and a double power law with an exponential cutoff, dPLE (5 free parameters: 2 normalizations, 2 indexes and the cutoff energy\footnote{For simplicity (i.e., to reduce the number of parameters) and since we expect the first population to be subdominant at high energy, we apply a single spectral cutoff.}). The analytical expressions of these two models are:
\begin{equation}
N_1\times(E_iE_j)^{-\alpha}e^{\left({-\frac{E_i+E_j}{E_{\rm{cut}}}}\right)}
\end{equation}
\begin{equation}
\left[N_1\times(E_iE_j)^{-\alpha}+N_2\times(E_iE_j)^{-\beta}\right]e^{\left({-\frac{E_i+E_j}{E_{\rm{cut}}}}\right)}
\end{equation}
The fit is performed on the C$_P^{ij}$ normalized by $E_i^2E_j^2/[(\Delta E_i)(\Delta E_j)]$, where $E_{i}$ and E$_{j}$ refer to the logarithmic center of the $i^{\rm{th}}$ and $j^{\rm{th}}$ energy bins, and the resulting best-fit parameters are summarized in Tab.~\ref{tab:fit_param}. 
The results of the best fits for the autocorrelation amplitudes $C_P$ are shown in Fig.~\ref{fig:Cp}.\\
The chi-square difference between the two best-fit configurations is $\Delta\chi^2 = \chi^2_{\rm{sPLE}} - \chi^2_{\rm{dPLE}} = 12.24$. In order to obtain the statistical significance of the result, we performed $10^7$ Monte Carlo samplings of the null hypothesis (the sPLE model) and derived the distribution of the chi-square differences, from which we determine a preference for the dPLE model at the $99.98\%$ CL (corresponding to $\sim3.7\sigma$). Details about the Monte Carlo can be found in the SOM.\\
The two power-law indices resulting from the best fit of the dPLE model are $-2.55\pm0.23$, for the low-energy component and $-1.86\pm0.15$, for the one dominating above a few GeV. \\
The best fit for the dPLE model reveals a transition range between the two populations around 4 GeV. Separating the first 4 energy bins from the following 6 bins (we exclude the last 2 energy bins, which are completely beyond $E_{{\rm cut}}$, in order to avoid energies affected by absorption by the extragalactic background light), we define 4 sub-rectangles of the cross-correlation coefficient matrix, and evaluate the mean and the standard deviation of the mean for each sub-rectangle.  The values are shown in the right-hand panel of Fig.~\ref{fig:rij}: the off-diagonal region deviates from 1 at $4 \sigma$, which unequivocally favors a double population scenario.\\
While detailed modeling of the underlying source classes is left for upcoming work, our findings  are compatible with most of the contributions being from blazar-like sources above a few GeV. \\
At lower energies, a population with a softer spectrum, such as possibly misaligned AGNs \citep{DiMauro:2013xta} or a different type of blazars \citep{2012ApJ...751..108A}, appears to dominate the UGRB. \\

\onecolumngrid

\begin{table}[t]
\centering
  \begin{tabular}{ c|c|c|c|c|c|c|c}
  \hline
  \hline
  \multicolumn{8}{c}{Fit Parameters} \\
  \hline
  Model & $N_1$ & $\alpha$ & $N_2$ & $\beta$ & $E_{{\rm cut}}$ & $\chi^2$ & DoF \\
  \hline
   sPLE & (2.7$\pm$0.3)E-18 & 0.13$\pm$0.03 &  -- & -- & 170$\pm$50 & 84.7 & 75 \\
 \hline 
   dPLE  & (3.5$\pm$0.8)E-18 & 0.55$\pm$0.23 & (7.6$\pm$6.4)E-19  & $-$0.14$\pm$0.15 & 89$\pm$24 & 72.5 & 73 \\
 \hline
 \hline
 \end{tabular}
\caption{Parameters of the fit of the global $C_P^{ij}$ energy spectrum for both a single power law with an exponential cutoff and for a double power law with an exponential cutoff. $E_{{\rm cut}}$ is in GeV, while $N_1$ and $N_2$ have the same dimension as $E^2$C$_P^{ij}$. DoF is the difference between the number of C$_P^{ij}$ considered ($(12\times(12+1))/2$) and the number of free parameters of the model. Since the fit has been performed on the C$_P^{ij}$ normalized by a factor whose global dimension is $E^2$, a factor of 2 should be added to the indices of the power laws to obtain the values in terms of intensity spectra.}
\label{tab:fit_param}
\end{table}

\twocolumngrid
\bigskip
\section{Acknowledgments}
\acknowledgments
The authors acknowledge Dr. Carlo Giunti for fruitful discussions about MC simulation procedures.The \textit{Fermi}-LAT Collaboration acknowledges support for LAT development, operation and data analysis from NASA and DOE (United States), CEA/Irfu and IN2P3/CNRS (France), ASI and INFN (Italy), MEXT, KEK, and JAXA (Japan), and the K.A.~Wallenberg Foundation, the Swedish Research Council and the National Space Board (Sweden). Science analysis support in the operations phase from INAF (Italy) and CNES (France) is also gratefully acknowledged. This work performed in part under DOE Contract DE-AC02-76SF00515. Some of the results in this paper have been derived using the HEALPix package. This work 
is supported by the ``Departments of Excellence 2018 - 2022'' Grant awarded by
the Italian Ministry of Education, University and Research (MIUR) (L. 232/2016).
MR acknowledges support by the Excellent Young PI Grant: ``The
Particle Dark-matter Quest in the Extragalactic Sky'' funded by the University of Torino and Compagnia di San Paolo and by ``Deciphering the high-energy sky via cross correlation'' funded by Accordo Attuativo ASI-INAF n. 2017-14-H.0. NF acknowledges the research grant ``The Anisotropic Dark
Universe" Number CSTO161409, funded under the program CSP-UNITO ``Research
for the Territory 2016" by Compagnia di Sanpaolo and University of Torino.

%

\onecolumngrid
\clearpage

\begin{center}
 \textbf{Supplemental Material: The unresolved gamma-ray sky through its angular power spectrum}
 \end{center}
\bigskip
\bigskip
\twocolumngrid
In the following sections we report further details about the correlation analyses. Specifically, in Sec.~I we report details about the construction of the masks, the procedure to subtract the Galactic emission, and the effect of this subtraction on the measurement; in Sec.~II we illustrate the computation of the window functions; in Sec.~III we outline the procedure to identify the optimal multipole range considered in the different energy bins when fitting the APS to compute the amplitude C$_P$ of the anisotropies; in Sec.~IV we discuss the noise term and the evidence of its underestimation when considering a Poissonian-only component, together with the definition of a method to overcome this problem; in Sec.~V we briefly comment on the study of the cross-correlations between macro energy bins and provide details about the Monte Carlo procedure to evaluate the significance of the result; in Sec.~VI we report and comment on comparison between our measurement and the previous obtained in Fornasa et al.; finally, in Sec. VII we illustrate two additional checks we performed to test the reliability of the mask deconvolution.

\section{I. Mask construction and \\foreground subtraction} \label{sec:Clean}
The mask consists of a region with shape determined by the intensity of the Galactic plane emission and circular regions around each source of the \FermiLAT 8-year (FL8Y) source list, which is a preliminary version of the 4FGL catalog; for energies above 10 GeV, we also include sources from the 3FHL catalog, which is derived specifically for this energy range. 

The Galactic plane mask, the same for all the energy bins, has been built by covering all the pixels where the flux of the Galactic model at 1 GeV is 3 times greater than the isotropic component evaluated at the same energy. The analysis is not sensitive to the exact value of this cut.

To build the mask of point sources, for each energy bin [$E\rm{_{min}}$, $E\rm{_{max}}$], we take the containment angle as given by the instrument point-spread function $\rm{PSF}(E_{\rm{min}})$\footnote{The PSF considered, according to our data selection, is the mean 68\% containment angle among PSF1, PSF2 and PSF3 type response functions.}, and we vary that value to define the radius $r_{src}$ of the covering region on each source, going from a minimum of 2$\times \rm{PSF}(E_{\rm{min}})$, for the faintest source (with flux\footnote{ We consider the integral photon flux from 1 to 100 GeV for FL8Y sources, and the corresponding quantity estimated from 10 GeV and 1 TeV for 3FHL sources.} $\phi_{\rm{min}}$), to a maximum of 5$\times \rm{PSF(E_{min})}$, for the brightest one (with flux $\phi_{\rm{max}}$); in formula:
$$\frac{r_{\rm{src}}(\phi_{\rm{src}}, E_{\rm{min}}) - 2 \times \rm{PSF}(E_{\rm{min}})}{5 \times \rm{PSF}(E_{\rm{min}}) - 2 \times \rm{PSF}(E_{\rm{min}})} = \frac{\log(\phi_{\rm{src}}) - \log(\phi_{\rm{min}})}{\log(\phi_{\rm{max}})-\log(\phi_{\rm{min}})}$$
 For bins above 14.5 GeV we conservatively keep E$\rm{_{min}}$=8.3 GeV, because the f$_{sky}$, namely the fraction of unmasked sky, is already greater than 55\%; hence we do not expect to gain much more statistics in further shrinking the mask. 
This procedure, performed separately for each energy bin considered in our anisotropy analysis, properly interpolates the improvement with energy of the PSF. Notice that the majority of the sources with very high flux are located near the Galactic plane, and are already covered by the Galactic plane mask. Extended sources are masked with large radii according to their size. One example of a mask is shown in Fig. 1 of the paper.

\begin{table}[t]
\centering
  \begin{tabular}{ c|c|c|c|c }
  \hline
  \hline
  \multicolumn{5}{c}{Parameters} \\
  \hline
  E$_{min}$ & E$_{max}$ & Masked catalog(s) & f$_{sky}$ & Counts \\
  $[{\rm GeV}]$ & $[{\rm GeV}]$ &  &  &  \\
  \hline
 0.5 & 1.0 & FL8Y & 0.134 & 577037 \\
 1.0 & 1.7 & FL8Y & 0.184 & 348514 \\
 1.7 & 2.8 & FL8Y & 0.398 & 321501 \\
 2.8 & 4.8 & FL8Y & 0.482 & 233035 \\
 4.8 & 8.3 & FL8Y & 0.549 & 120476 \\
 8.3 & 14.5 & FL8Y & 0.574 & 56535 \\
 14.5 & 22.9 & FL8Y+3FHL & 0.574 & 21399 \\
 22.9 & 39.8 & FL8Y+3FHL & 0.574 & 12464 \\
 39.8 & 69.2 & FL8Y+3FHL & 0.574 & 5159 \\
 69.2 & 120.2 & FL8Y+3FHL & 0.574 & 1955 \\
 120.2 & 331.1 & 3FHL & 0.597 & 1013 \\
 331.1 & 1000.0 & 3FHL & 0.597 & 150 \\
 \hline
 \hline
 \end{tabular}
\caption{Main parameters of our analysis for each energy bin: energy boundaries, E$_{min}$ and E$_{max}$; the \textit{Fermi} source catalog used to build the mask; the fraction of the sky outside the mask, f$_{sky}$; the number of photons outside the mask.}
\label{tab:Param}
\end{table}

\begin{figure}[h]
\centering
\includegraphics[width=8cm]{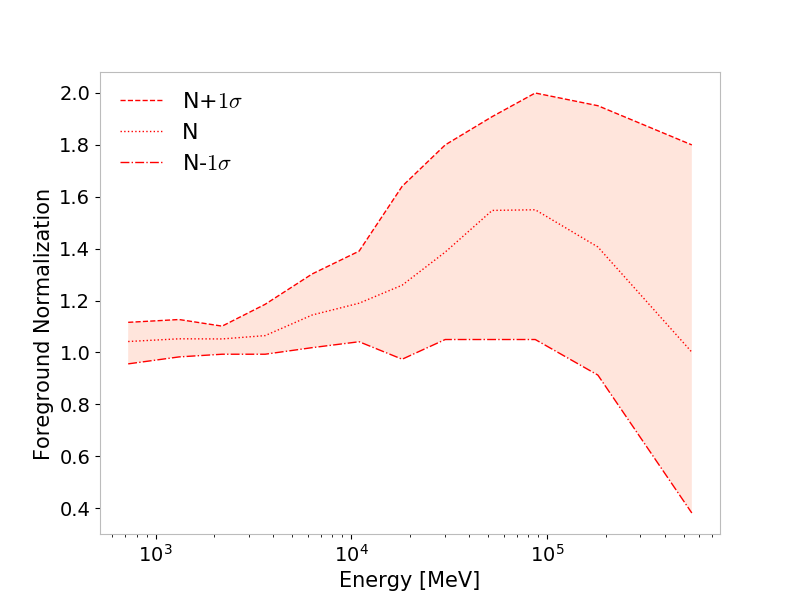}
\includegraphics[width=8cm]{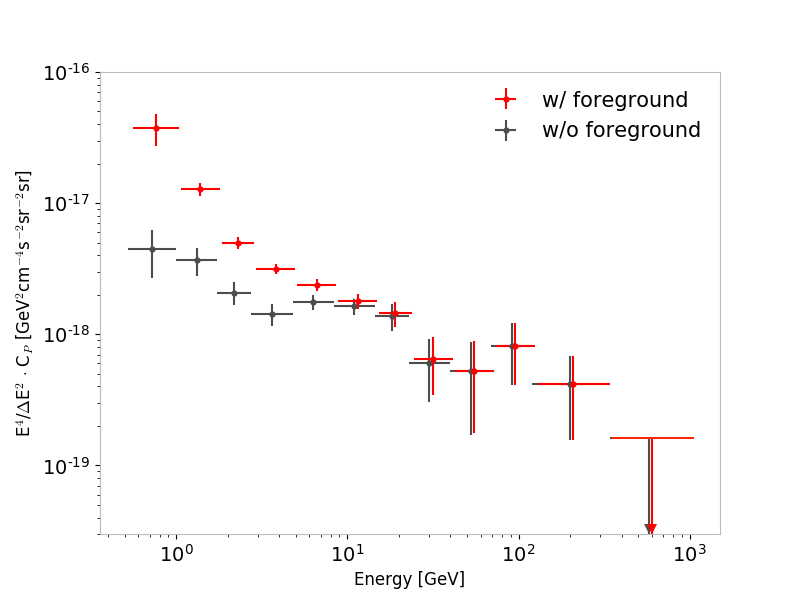}
\caption{Left: Normalization factor resulting from the Poissonian fit of the Galactic foreground to the data outside the mask; for visualization purposes, we report the normalizations for the macro energy bins computed averaging the values of the micro ones. Right: autocorrelation anisotropy energy spectrum with and without foreground subtraction. In both cases monopole and dipole terms have been removed from intensity maps prior to the APS computation.}
\label{fig:foresys}
\end{figure}

Once the Galactic foreground and individual sources mask is applied, we implement a further Galactic emission subtraction. This is performed in each micro energy bin of our analysis with the following procedure: assuming that the detected photons in each pixel outside the mask are the sum of an isotropic emission component $C$ and a high-latitude Galactic diffuse emission, we adopt the Galactic diffuse emission model gll\_iem\_v6.fits described in \citep{Acero:2016qlg} and we fit it to data leaving free its normalization factor $N$ and the constant term $C$. This maximum likelihood fit for Poissonian statistics is performed over the micro energy binned data, and is done with maps downgraded to order 6 to limit the effects of statistical fluctuations of the counts in each pixel. The resulting normalization factors $N$ are compatible with 1 at every energy and are shown in Fig.~\ref{fig:foresys} (left panel). They are then applied to the order 9 model maps, which are then subtracted from the corresponding order 9 data maps.

The effect of the foreground, even if not subtracted, is negligible at energies above $\sim$10 GeV, and this is shown in Fig.~\ref{fig:foresys} (right panel), where we compare the anisotropy energy spectrum obtained with and without foreground subtraction. Several effects contribute to the explanation why the Galactic emission is not significantly affecting the measurement, mostly at high energy, is due to several combined effects: 1) the foreground mask itself is efficient, 2) we consider only multipoles above $\ell=50$, and 3) the energy spectrum of the foreground is significantly softer than the UGRB spectrum above a few GeV (and before the high-energy cutoff). We tested the possible systematics of this subtraction procedure by repeating the measurement considering as normalization factor of the model the upper (N+$1\sigma$) or the lower (N-$1\sigma$) values of the N factor uncertainty band (reported as a shaded region in the left panel of Fig.~\ref{fig:foresys}); this test did not highlight any evident systematics.

One issue related to the the foreground model regards the resolution of the interstellar gas component measurements at high latitudes ($\sim$ 0.7$^{\circ}$) which is adopted in the public model gll\_iem\_v6.fits: this is worse than the instrument PSF for energies above 1 GeV. As we know that the Galactic foreground is not affecting our measurements for energies above $\sim$10 GeV, the coarse-grained resolution of the gas-correlated component would potentially affect our foreground-subtraction in our second, third, fourth, and fifth energy bins (those between 1 and 10 GeV). We verified that using higher resolution maps (under internal development) does not affect our results.

We do not model the solar inverse Compton emission: it is localized along the ecliptic and hence expected to only contribute at very low multipoles, which we exclude from our analysis. Our case is analog to that discussed in Fornasa et al., and we refer to Sec. D of that work, which demonstrates that the effect of this component on the final anisotropy energy spectrum is negligible.

Any leakage from resolved point-like sources due to PSF effects could affect the APS measurement. We tested the reliability of our masks by considering an additional term in the Poissonian fitting procedure to extract the UGRB signal: a resolved point-like source template based on the FL8Y source list. To build such a template we considered the spectral shape and parameters provided with the FL8Y list for each source and then we integrated the resulting spectrum between $E_{min}$ and $E_{max}$ to compute the total flux in a given energy bin [$E_{min}$--$E_{max}$]. 
We created point source maps for each energy interval by placing these fluxes at relative source locations and then convolving with the instrumental PSF($\theta$).
The $\Delta$(ln(Likelihood)) for each energy bin does not reveal a preference for the model with the additional source template: this test confirms that we have no leakage from point-like source emission outside our masks.

\section{II. Window functions} \label{sec:Wfunc}

\begin{figure}[t]
\centering
\includegraphics[width=8cm]{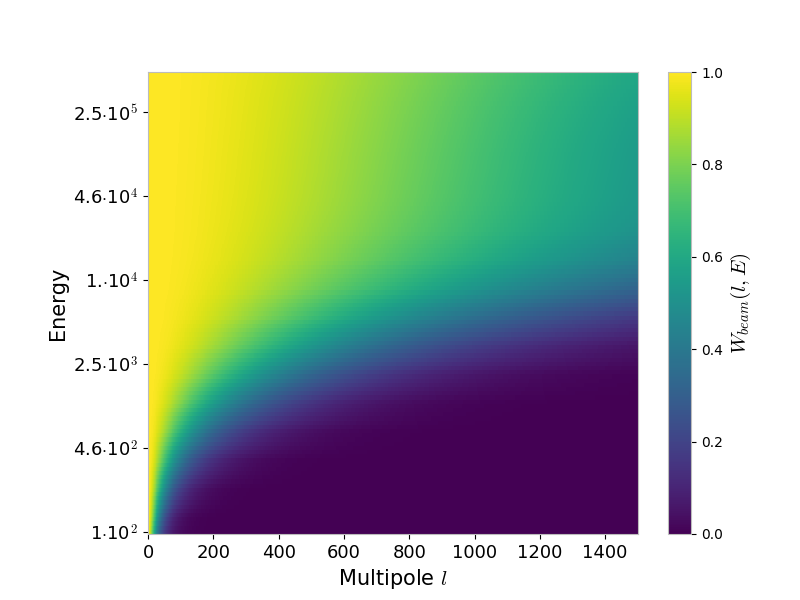}
\includegraphics[width=8cm]{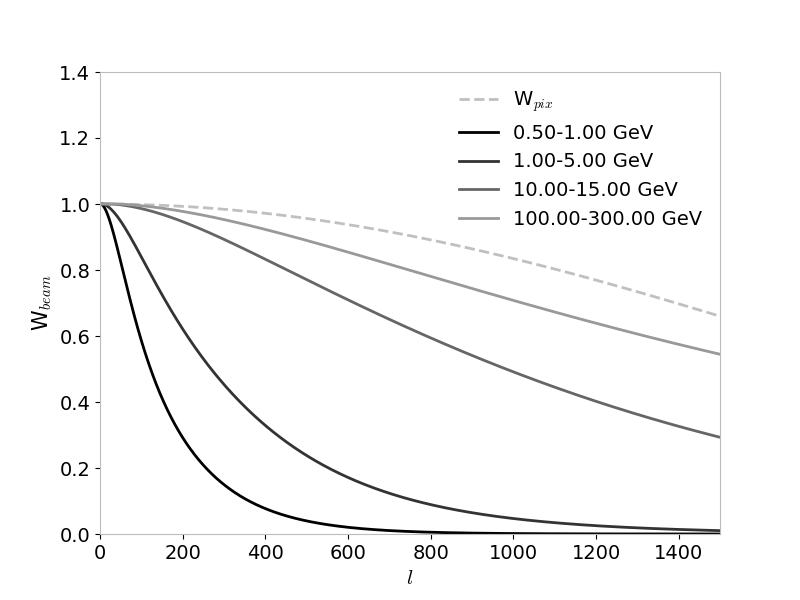}
\caption{Left: Beam window function W$_{beam}$ as a function of energy and the multipole for the SOURCEVETO PSF1+2+3 event types; this function has been computed for 100 energies between 100 MeV and 1 TeV and for all the multipoles from 0 to 1500. Right: Pixel W$_{pix}$ and beam W$_{beam}$ window functions as a function of the multipole $\ell$ and averaged in different energy bins as described in Sec. III.}
\label{fig:Wfunction}
\end{figure}

The map resolution and the effect of the PSF are taken into account by the pixel window function, $W^{pix}(\ell)$, and by the beam window function, $W^{beam}(E,\ell)$, respectively. The former is computed using the HEALPix routine \texttt{pixwin}, while the latter is determined as:

$$W^{beam}(E,\ell) = 2\pi \int_{0}^{\pi} P_{\ell}(\cos\theta) {\rm PSF}(\theta,E) \, \sin\theta d\theta$$
where $P_{\ell}(\cos\theta)$ are the Legendre polynomials of index $\ell$. Since the PSF is a function of energy, it is necessary to average $W^{beam}(E,\ell)$ inside each energy bin weighted by the intensity energy spectrum of the UGRB, which is approximately a power law with index $-2.3$ \citep{2015IGRB}. This gives the bin-averaged beam function $W^{beam}_E(\ell)$, where E here labels the energy bin under consideration. 
Fig.~\ref{fig:Wfunction} shows $W^{beam}(E,\ell)$ as a function of energy and the multipole for the SOURCEVETO PSF1+2+3 class of events and some of its averages $W^{beam}_E(\ell)$.

We call $W_{E}(\ell)$ the product of the bin-averaged beam window function and the pixel window function.

\section{III. From APS to C$_P$} \label{sec:aps_fit}
Fig.~\ref{fig:AutoAPSfromecross} shows the APS for all the energy bins: they are compatible with a flat spectrum, as expected if the anisotropy signal is dominated by a population of unresolved point-like sources isotropically distributed in the sky. We binned the obtained raw APS from PolSpice, as well as the associated covariance matrix returned by the algorithm, into 26 logarithmic multipole bins in order to smooth the intrinsic fluctuation of this function; to do this we implemented the unweighted averaging procedure proposed in Fornasa et al., which was validated with Monte Carlo simulations (see Sec. IV-A of Fornasa et al.). Assuming a signal dominated by the correlation at zero angular separation of an unresolved population (see Introduction), we fit the APS with a constant in a specific range of multipoles (as defined below): this returns the level of anisotropy $C_P$ in each energy bin. Since at the lowest multipoles a residual Galactic foreground contamination might be present, while at the highest multipoles the correction for the PSF can introduce inaccuracies, a proper definition of an energy-dependent range of multipoles in which to perform the fit is crucial. 

\begin{figure}[h]
\centering
\includegraphics[width=8cm]{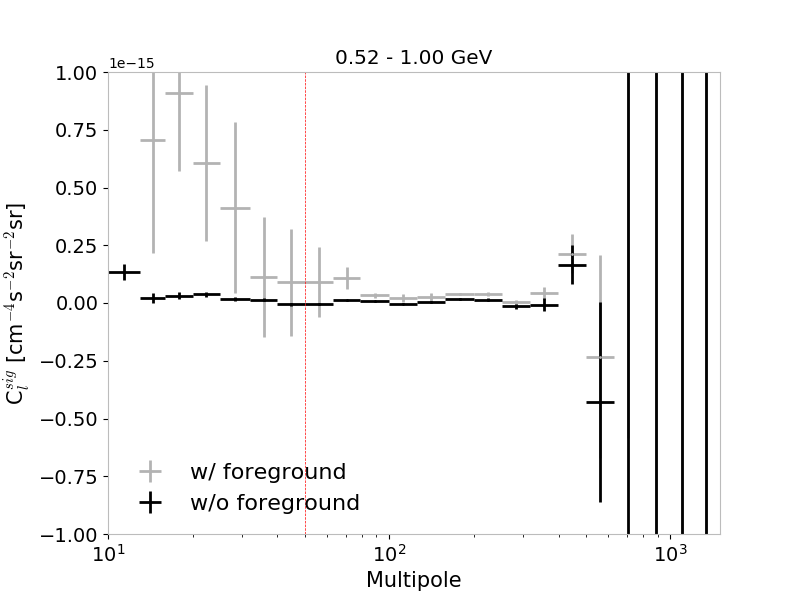}
\includegraphics[width=8cm]{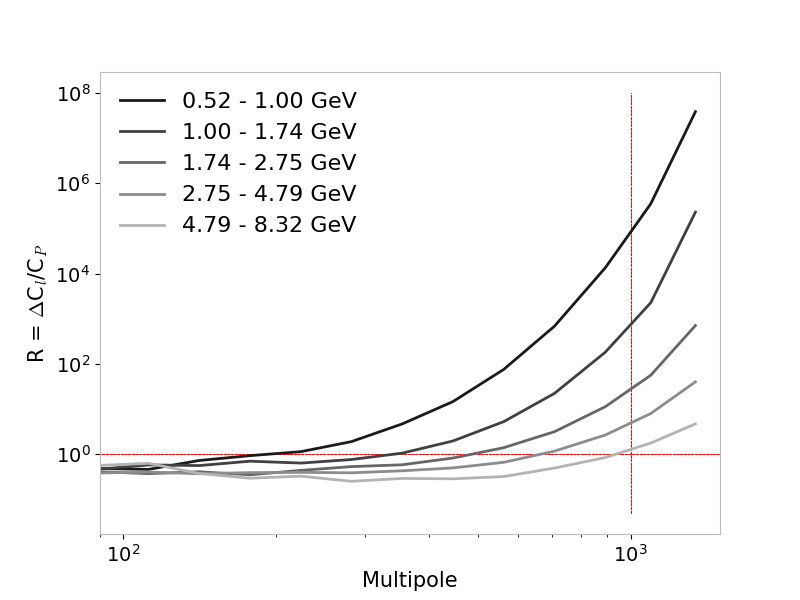}
\caption{Left: APS with and without Galactic foreground subtraction for the first energy bin ($0.5-1.0$) GeV. The red line indicates the lower multipole value considered in the APS fit. Right: the trust function $R$ as a function of multipole for various energy bins: $(0.5-1.0)$ GeV, $(1.0-1.7)$ GeV, $(1.7-2.8)$ GeV, $(2.8-4.8)$ GeV, $(4.8-8.3)$ GeV, for which the method leads to  $l_{max}=150$, $l_{max}=250$, $l_{max}=450$, $l_{max}=600$, and $l_{max}=900$ respectively. $l_{max}=1000$ is associated with all the remaining energy bins.}
\label{fig:Cls_ex}
\end{figure}

Regarding the large physical scales (low multipoles), we apply a cut for  multipoles below $\ell_{min} = 50$ at each energy. Fig.~\ref{fig:Cls_ex} shows the case of the first energy bin, where the Galactic foreground is expected to contribute the most: the effect of the foreground on the APS is in fact relevant at low multipoles: it affects the measurement and contributes to considerably increase the uncertainty below $\ell = 50$, while its effect is reduced above this threshold. Fig.~\ref{fig:Cls_ex} also shows that subtraction is quite effective in reducing the impact of Galactic foreground in a wide multipole range. Even though the effect of Galactic foreground on the APS decreases with energy, to stay on the safe side we apply a cut $\ell_{min} = 50$ at all energies.

To determine the upper extent on the multipole range for the fit, we define a fiducial test which is sensitive to both the angular (and therefore $\ell$) extent of the PSF and the reduction of statistics at increasing energies. We define a trust function\footnote{The influence of the PSF correction on the $C_P$ can be found in the same definition of the APS, while for the case of the $C_{\ell}$ error it can be understood by looking at its analytical definition: $\Delta C_{\ell}=\sqrt{\frac{2}{2\ell+1}} \bigg(C_{\ell} + \frac{C_N}{W_{\ell}^2}\bigg)$} $R(l) = \Delta C_{\ell}/C_P$ as the ratio between the absolute error of the APS at any given $\ell$, and the $C_P$ obtained by fitting the APS {\sl up to} that multipole $\ell$. The maximal multipole $\ell_{max}$ up to which we trust the PSF correction is the one at which $R(\ell)$ becomes greater than 1. An example of this procedure is shown in the right panel of Fig.~\ref{fig:Cls_ex}. In any case the maximum value considered is 1000, even if our method would allow a greater $l_{max}$, in order to avoid any edge effects related to the APS computation. The gray regions in the plots of Fig.~\ref{fig:AutoAPSfromecross} illustrate the range of multipoles in which the APS is fitted: one can notice how $\ell_{max}$ increases to higher multipoles as the energy increases, due to the progressive improvement of the PSF as a function of energy.
The APS fit is performed considering the covariance matrix returned by PolSpice, for the range of multipoles determined by the procedure described above.

\section{IV. Estimation of the Poissonian noise} \label{sec:noise}

This section reviews the standard procedure to compute the autocorrelation, focusing on the importance of the removal of the noise component, and motivates the need to define the new technique presented in Sec. III of the paper  in order to obtain the autocorrelation APS.

The Poissonian noise $C_N$ is a constant term in the APS, and can be estimated as $C_N = \frac{\langle n^i_{\gamma,{\rm pix}}/(\epsilon^i_{{\rm pix}})^2 \rangle} {\Omega_{{\rm pix}}}$. Being a shot noise term, it affects only  the determination of the autocorrelation. In order to interpolate the variation in the detector exposure, it is computed in each ``micro" energy bin and then summed in order to obtain the value for the ``macro" energy bins for which we present our analysis. The C$_N$ values for each ``macro" bin are reported in Tab.\ref{tab:Cn-param}.

The autocorrelation APS returned by PolSpice must be corrected for this noise: this is achieved by subtracting the C$_N$ constant term, after which the correction due to the PSF and pixel effects are applied:
\begin{equation}
C_{\ell} = \frac{C_{\ell}^{Pol}-C_N}{W_{E}^2(\ell)}
\label{eq:subtr}
\end{equation}
Let us call this the {\sl standard} method.

For each APS in each macro energy bin we then compute the C$_P$ values and we obtain the anisotropy energy spectrum shown in Fig.~\ref{fig:Cps_auto}, top-left panel. In the top-right panel we also show the cross-correlation between two independent data samples: the first 4 years (F4yrs) against the second 4 years (L4yrs) of data (i.e., the result of correlating F4yrs $\times$ L4yrs). The two samples have been subject to the same data selection discussed in Sec. II of the paper for the full analysis. The advantage in performing this check is that random fluctuations do not correlate between the two separate time periods, and a white noise subtraction is no longer needed for this cross-in-time correlation. This result is a {\sl noiseless} realization of the anisotropy study, but being based on half the statistical sample it is affected by a larger error. If the noise estimate $C_N$ is correct, its subtraction in Eq.~\ref{eq:subtr} is proper and the results obtained by the {\sl standard} and {\sl noiseless} determination of the C$_P$ should coincide. The top-right panel in Fig.~\ref{fig:Cps_auto} shows that this appears not to be the case, and suggest that a misestimation of the noise is present.

To investigate more deeply this point, in Fig.~\ref{fig:Cps_auto} (top-right panel) we also report the autocorrelations of the two sub-samples, F4yrs and L4yrs (i.e. F4yrs $\times$ F4yrs and L4yrs $\times$ L4yrs). We notice that, above a few GeV, the cross-correlation spectrum is systematically below the autocorrelation one, while the autocorrelations for the two smaller data samples are systematically above. 

Since the total data sample is in fact composed of the union of F4yrs and L4yrs samples, which are independent, it should be that: 
\begin{equation}
C_P = \frac{1}{4}(C_P^{1\times 1} + C_P^{2\times 2} + 2C_P^{1\times 2})
\end{equation}
where the indexes 1 and 2 refer to F4yrs and L4yrs respectively.
Indeed, this is what we obtain, as illustrated in the bottom-left panel of Fig.~\ref{fig:Cps_auto}: this proves the consistency of all the measurements among themselves and points to an underestimation of the random noise $C_N$ in the autocorrelations. This effect appears larger when the statistics are reduced (as occurs when we consider the two time bin samples, each of which has roughly half of the statistics of our full analysis). We find the same result (increase of the autocorrelation APS when the statistical sample is reduced) when we define the time bins as the first four even years (E4yrs, namely the 2$^{\rm{nd}}$, 4$^{\rm{th}}$, 6$^{\rm{th}}$, 8$^{\rm{th}}$) and the first four odd years (O4yrs, namely the 1$^{\rm{st}}$, 3$^{\rm{rd}}$, 5$^{\rm{th}}$, 7$^{\rm{th}}$); this should exclude systematics related to the time bins definition or unaccounted transients. 

To confirm that the underestimation of the noise is accentuated when reducing the statistics of the sample, we applied our analysis to the PSF3 type events (which represent 1/3 of the data we use by adopting the PSF1+2+3 event types). In this case as well we find an autocorrelation spectrum above the one obtained with the whole sample. The same effect has been observed also when reducing the dimension of the energy bins, and an opposite behavior is found when enlarging the energy binning. 

All this evidence suggests that the estimation of the noise in our measurement is not effective. 
A possible explanation could relate to a small deviation of the random noise distribution from a Poissonian, which could be induced, e.g., by data cleaning procedures or to an inaccurate estimator of the Poissonian noise (to estimate the noise, we use measured counts in each pixel which are just an estimator of the true photon emission).
Another possible explanation is related to the effectiveness of the mask removal algorithm in PolSpice: it can introduce a small bias in the autocorrelation APS of noise-dominated maps (the cross-correlation case is exempt from such a bias, since it does not exhibit noise).
This additional unaccounted for effect is of the order of one percent of the estimated Poissonian C$_N$ term that we estimate. 
Since the APS signal is largely sub-dominant at high energy as compared to the noise (which makes noise subtraction not only necessary but also delicate), it is clear that even a small deviation of the order of a few percent of the subtracted C$_N$ can result in a different measurement of the signal up to a $10\%$ (this is the case of the highest energy bin, where the estimated Poissonian noise is two orders of magnitude higher than the C$_P$ value), as can be seen by comparing the various estimates reported in Tab.~\ref{tab:Cp-param} and Tab.\ref{tab:Cn-param}.

For this reason we decided to adopt a different technique to measure the autocorrelation APS in the various (macro) energy bins, that for definiteness we call {\sl improved noiseless}. The technique is outlined in the main section of the paper, and is based on the fact that if the sources contributing to the APS signal have broad energy spectra (with no structure on the scale of our micro energy bins, as is expected for any astrophysical source), we can assume that the autocorrelation spectra of the micro bins are well approximated by the average APS of all the cross-correlations between the same bins. In this case the APS is simply given by Eq. 2 of the paper.

In the bottom-right panel of Fig.~\ref{fig:Cps_auto} we show the two measurements of anisotropy energy spectrum, obtained with the {\sl standard} analysis and with the {\sl noiseless improved} one, and we also report the statistical error band associated to the estimator of the noise term, applied to the standard measurement only. Tab.~\ref{tab:Cn-param}  outlines the impact of the unaccounted for noise term and the size. The C$_P$ values obtained with the {\sl standard} analysis are reported in Tab.~\ref{tab:Cp-param}.

The APS estimator in Eq. 2 also could be computed by approximating each $C_{\ell,\rm{micro}}^{\delta\delta}$ term with a weighted mean of all the cross-correlation terms $C_{\ell,\rm{micro}}^{\gamma\delta}$\footnote{Here we use $\gamma$ and $\delta$ in place of $\alpha$ and $\beta$ adopted in Eq. 2 in order to not generate confusion between the energy micro binning nomenclature and the fit parameters in Tab. II.} (in place of the arithmetic mean adopted in the current work), the weights being $(E_{\gamma}E_{\delta})^{-\alpha}$, where $\alpha$ is taken from the best fit of the sPLE model. This second estimator of the autocorrelation in each macro bin returns a $C_P$ value almost identical to the one presented in this work.

\begin{table}[t]
\centering
  \begin{tabular}{ c|c|c }
  \hline
  \hline
  \multicolumn{3}{c}{Anisotropy energy spectrum parameters } \\
  \multicolumn{3}{c}{for standard autocorrelation analysis} \\
  \hline
  E$_{min}$& E$_{max}$& C$_P\pm\delta$C$_P$ \\
  ~&~&~\\
  $[{\rm GeV}]$ & $[{\rm GeV}]$ & [${\rm cm}^{-4}{\rm s}^{-2}{\rm sr}^{-2}{\rm sr}$]  \\
  \hline
 0.5 & 1.0 & (4.9$\pm$1.6) E-18 \\
 1.0 & 1.7 & (6.3$\pm$1.6) E-19 \\
 1.7 & 2.8 & (1.2$\pm$0.2) E-19 \\
 2.8 & 4.8 & (4.7$\pm$0.8) E-20 \\
 4.8 & 8.3 & (2.0$\pm$0.2) E-20 \\
 8.3 & 14.5 & (6.0$\pm$0.8) E-21 \\
 14.5 & 22.9 & (1.7$\pm$0.3) E-21 \\
 22.9 & 39.8 & (6.0$\pm$1.3) E-22 \\
 39.8 & 69.2 & (1.7$\pm$0.5) E-22 \\
 69.2 & 120.2 & (6.9$\pm$1.9) E-23 \\
 120.2 & 331.1 & (2.3$\pm$1.0) E-23 \\
 331.1 & 1000.0 & (1.7$\pm$1.5) E-24 \\
 \hline
 \hline
 \end{tabular}
\caption{Values of the measured autocorrelation amplitudes C$_P$ and their corresponding errors $\delta$C$_P$ for each energy bin, as obtained from the {\sl standard} autocorrelation analysis.}
\label{tab:Cp-param}
\end{table}

\begin{table}[t]
\centering
  \begin{tabular}{ c|c|c|c|c|c }
  \hline
  \hline
  \multicolumn{6}{c}{Relevant parameters for the noise contribution study} \\
  \hline
 E$_{min}$ & E$_{max} $ & C$_N$ & $\frac{\delta C_{N}}{C_N}$ & $\frac{|C_P^{std} - C_P^{imp}|}{C_N}$ & $\frac{C_{P}^{imp}}{C_N}$ \\
~&~&~&~&~&~\\
$[{\rm GeV}]$ & $[{\rm GeV}]$ & [cm$^{-4}$s$^{-2}$sr$^{-1}$] & $[\%]$ & $[\%]$ & $[\%]$ \\
  \hline
0.5 & 1.0 & 1.056 E-17 & 0.132 & 11.995 & 34.647 \\
1.0 & 1.7 & 3.548 E-18 & 0.169 & 1.544 & 17.282 \\
1.7 & 2.8 & 1.375 E-18 & 0.176 & 1.915 & 6.825 \\
2.8 & 4.8 & 8.324 E-19 & 0.207 & 1.627 & 4.074 \\
4.8 & 8.3 & 3.904 E-19 & 0.288 & 1.444 & 3.566 \\
8.3 & 14.5 & 1.768 E-19 & 0.421 & 0.983 & 2.409 \\
14.5 & 22.9 & 6.899 E-20 & 0.684 & 1.215 & 1.301 \\
22.9 & 39.8 & 3.895 E-20 & 0.896 & 1.008 & 0.537 \\
39.8 & 69.2 & 1.576 E-20 & 1.392 & 0.672 & 0.376 \\
69.2 & 120.2 & 6.205 E-21 & 2.262 & 0.608 & 0.497 \\
120.2 & 331.1 & 3.287 E-21 & 3.142 & 0.346 & 0.356 \\
331.1 & 1000.0 & 5.094 E-22 & 8.165 & 0.433 & -0.086 \\
 \hline
 \hline
 \end{tabular}
\caption{For each energy bin, we show the Poissonian noise term C$_N$ and its relative statistical error $\delta C_{N}/{C_N} = 1/\sqrt{N}$, $N$ being the number of counts in that macro energy bin; we then report the fraction of ``missing noise'' (relative to the Poissonian noise) arising from the determination of the autocorrelation amplitude 
$C_P$ in the given macro energy bin with the {\sl standard} autocorrelation analysis ($C_P^{std}$) and the {\sl improved noiseless} $C_P^{imp}$, obtained with the technique of determining the autocorrelation in the macro energy bin by using the information from the cross-correlations between the micro energy-bins that form the macro energy bin (see text for details). The last column shows the ratio between the $C_P^{imp}$ and the Poissonian noise $C_N$.}
\label{tab:Cn-param}
\end{table}

\section{V. Cross APS} \label{sec:Miscell}

The cross-APS between intensity maps in two different macro energy bins, $E_i$ and $E_j$, is computed as:
\begin{equation}
C_\ell^{ij} = \frac{C_\ell^{ij,Pol}}{W_{E_i}(\ell)W_{E_j}(\ell)}
\end{equation}
In analogy to the determination of the autocorrelation APS, the amplitudes of the cross C$_P^{ij}$ are determined by fitting the measured $C_\ell^{ij}$ with a constant value. For each couple of cross-correlated energy bins, the range of multipoles considered to compute the C$_P^{ij}$ (the interval between $\ell_{min}=50$ and the energy-dependent $\ell_{max}$) is the smaller of the two intervals derived for $E_i$ and $E_j$ with the method of the trust function $R$ discussed in Sec. III and shown in Fig.~\ref{fig:AutoAPSfromecross} (i.e., it is the interval pertaining to the lowest energy between $E_i$ and $E_j$ ). In Tab.~\ref{tab:CpCross_val} we report the values of the C$_P^{ij}$ for each pair of $E_i$ and $E_j$ bins.

Regarding the analysis of the auto- and cross-correlations aimed at determining the underlying field of sources, we perform a 2-dimensional fit as outlined in Sec. IV of the main paper. In Fig~\ref{fig:cpij_fit} the three panels report examples of cross-correlation anisotropy energy spectra for three different macro energy bins correlated with all the others, and the best-fit curves at those values of $E_i$ and $E_j$ are also reported.

In the determination of the $\chi^2$, we neglect covariance between signals in different macro energy bins. This is justified by the fact that the main source of error is given by the photon noise, which is uncorrelated in different energy bins.\\

The chi-square difference between the two best-fit configurations is $\Delta\chi^2 = \chi^2_{sPLE} - \chi^2_{dPLE}$ = 12.24. In order to obtain the statistical significance of the result, we performed a Monte Carlo simulation to compute the $\Delta\chi^2$ distribution in the assumption that the null hypothesis (sPLE model) holds true. We produced $10^7$ fake-data samples generating fake $C_P^{ij}$values for each $E_iE_j$  from the best-fit parameters of the sPLE model, and randomized them according to gaussian distributions with standard deviations equal to the statistical errors of real data. Each fake-data sample is then fitted with both sPLE and dPLE models, the $\Delta\chi^2$ is computed, and eventually we obtain the $\Delta\chi^2$ distribution. The latter provides information about the $CL$ of rejection of the null hypothesis: in particular we find a $p$-value $=0.0002$ and determined that for real data sPLE model is rejected at the $99.98\%~CL$ (corresponding to $\sim3.7\sigma$). In Fig.~\ref{fig:cpij_fit} (bottom-right) we report the $\Delta\chi^2$ distribution obtained from this Monte Carlo procedure compared to a $\chi^2$ distribution with 2 degrees of freedom, which is the one predicted by Wilks' theorem \citep{Wilks:1938} for the case of linear and nested models\footnote{Due to the non-linearity (in the parameters) of our models the applicability of Wilks' theorem is not guaranteed and a Monte Carlo simulation is required.}.

\section{VI. Comparison with previous measurement} \label{sec:CuocoComp}
\begin{table}[h]
\centering
  \begin{tabular}{ c|c|c|c }
  \hline
  \hline
  \multicolumn{4}{c}{Parameters} \\
  \hline
 E$_{min}$ & E$_{max}$ & f$_{sky}$ & C$_P\pm\delta $C$_P$ \\
 $[{\rm GeV}]$ & $[{\rm GeV}]$ & & [${\rm cm}^{-4}{\rm s}^{-2}{\rm sr}^{-2}{\rm sr}$] \\
  \hline
 0.5 & 1.0 & 0.199 & (1.5$\pm$0.3) E-17 \\
 1.0 & 1.7 & 0.250 & (1.8$\pm$0.2) E-18 \\
 1.7 & 2.8 & 0.443 & (3.3$\pm$0.2) E-19 \\
 2.8 & 4.8 & 0.511 & (1.4$\pm$0.08) E-19 \\
 4.8 & 8.3 & 0.564 & (4.1$\pm$0.2) E-20 \\
 8.3 & 14.5 & 0.586 & (1.2$\pm$0.08) E-20 \\
 14.5 & 22.9 & 0.586 & (2.9$\pm$0.3) E-21 \\
 22.9 & 39.8 & 0.586 & (1.3$\pm$0.1) E-21 \\
 39.8 & 69.2 & 0.586 & (3.6$\pm$0.5) E-22 \\
 69.2 & 120.2 & 0.586 & (9.1$\pm$2.0) E-23 \\
 120.2 & 331.1 & 0.586 & (3.6$\pm$1.0) E-23 \\
 331.1 & 1000.0 & 0.586 & (2.9$\pm$1.5) E-24 \\
 \hline
 \hline
 \end{tabular}
\caption{The sky coverage fraction f$_{sky}$, the anisotropy amplitudes C$_P$, and their relative errors $\delta$C$_P$ for each energy bin in the case the 3FGL catalog is used to mask the sky and the {\sl standard} analysis is adopted. The ranges of multipoles considered in the fit of the APS are the same used for the nominal case and reported in Tab. I of the paper.}
\label{tab:Cp-param3FGL}
\end{table}

Fig.~\ref{fig:cp3FGL} compares the measurement of the anisotropy energy spectrum obtained in this paper (with point sources masked according to the FL8Y source list, and with the adoption of our {\sl improved} technique) with the measurement we obtain by adopting the 3FGL catalog for masking resolved point sources and by using the  {\sl standard} analysis technique. The latter can be directly compared to the analysis of Fornasa et al., which adopted the same 3FGL catalog and the same noise-subtraction technique, but with lower photon count statistics.

We note that the results obtained with the 3FGL catalog, reported in Tab.~\ref{tab:Cp-param3FGL}, are totally consistent with the results of the previous investigation, the only difference being the gain in statistics which manifests itself in a noticeable reduction of the error bars. The greater statistics here are possible due to a larger integration time for the detector and also thanks to an improved technique adopted to define the point source mask.

The comparison with the new measurement performed by masking the FL8Y source list shows the extent of the reduction of the APS as a consequence of the reduced number of unresolved sources left after FL8Y. Since the anisotropy energy spectrum found masking the 3FGL sources shows similar features\footnote{Best-fit parameters for dPLE model when considering 3FGL catalog are: $N_1=(1.30\pm0.03)10\times^{-17}$, $\alpha=0.45\pm0.05$, $N_2=(8.0\pm2.9)\times10^{-19}$, $\beta=-0.29\pm0.06$ and $E_{{\rm cut}}=94\pm8$ (parameters units are the same declared in the caption of Tab. II of the paper).} to the result presented here, we expect the low-energy population to be present in the FL8Y catalog and to be a relevant fraction of objects not associated with any source of the 3FGL catalog.

\section{VII. Test on mask deconvolution}
\label{sec:masktest}
In order to check that mask deconvolution does not introduce artifacts in the measured angular power spectrum, we performed two additional checks. In the first, we have have performed an apodization of our masks with a scale of the order of half the masking radius defined for each source, adopting a sine behavior for the apodizing function. This operation should prevent possible artifacts driven by the sharp edges of the masked regions. Fig.~\ref{fig:cpdiff_sim} (left panel) shows that the results obtained with apodization are well consistent with the results obtained with our nominal masks, while exhibiting a larger error since the fraction of the unmasked sky is reduced.

The second check was an evaluation of the possible effects induced by the mask through simulated full-sky gamma-ray maps. We generated $10^4$ maps with the same (2-point) statistical correlation as the one estimated for the real sky in our analysis. We adopted the {\tt synfast} generator provided by HEALPix, starting from a flat (in multipole) angular power spectrum with amplitude equal to our measurement. Each generated map was then analyzed with the PolSpice algorithm in two ways: first without a mask and then with the same mask adopted in our analysis. The distribution of the relative differences between the reconstructed $C_P$ from the full-sky and masked-sky analyses is shown in Fig.~\ref{fig:cpdiff_sim} (right panel) for two energy bins, $(1.0 - 1.7)$ GeV and $(8.3 - 14.5)$ GeV. The shaded bands indicate the corresponding statistical relative errors we obtained in the nominal analysis. We note that the differences are distributed around zero, with a width much smaller than the statistical error of our measurement. This demonstrates that the mask deconvolution is performed properly and it does not translate into a bias of the measured $C_P$.

\onecolumngrid

\begin{addmargin}[-2em]{-2em}
\noindent
\begin{figure}[h]
\includegraphics[width=5.5cm]{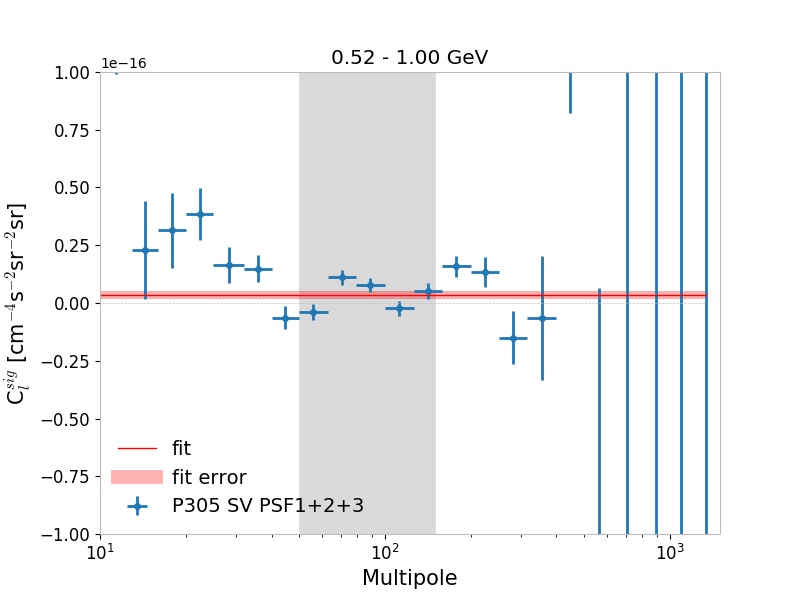}
\includegraphics[width=5.5cm]{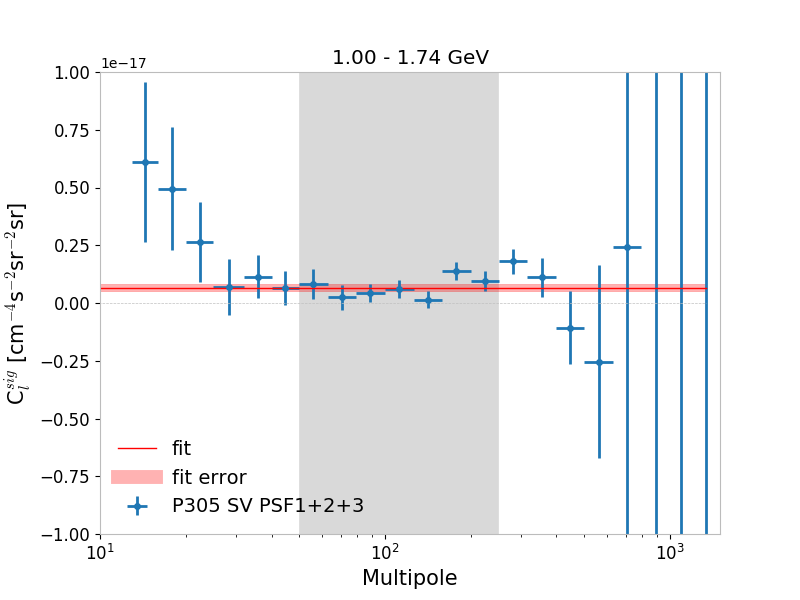}
\includegraphics[width=5.5cm]{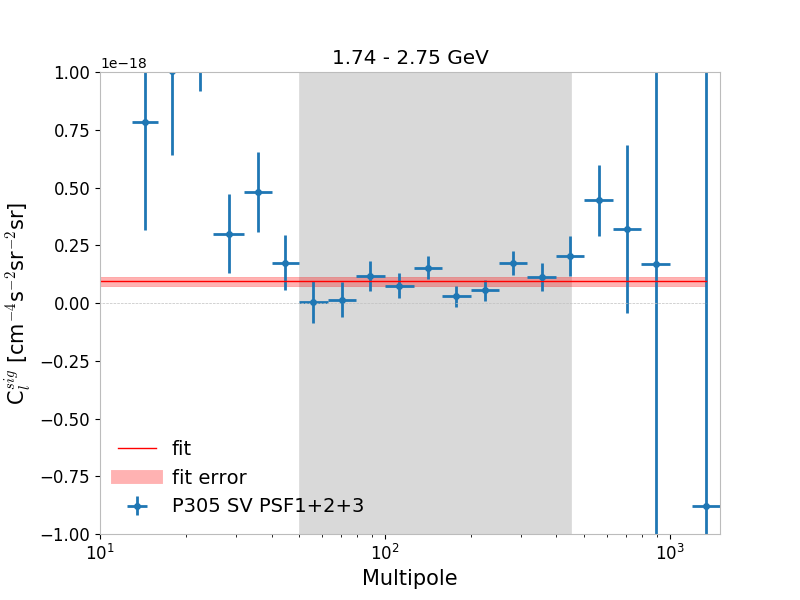}
\includegraphics[width=5.5cm]{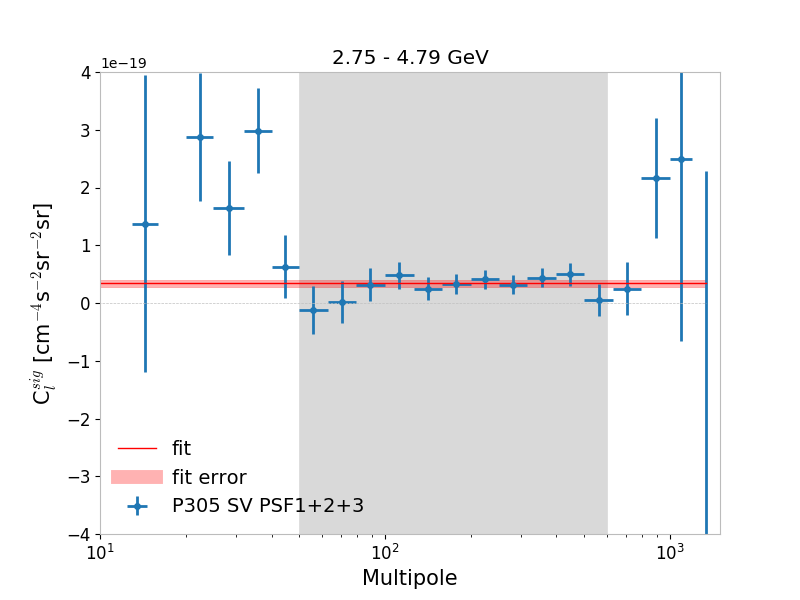}
\includegraphics[width=5.5cm]{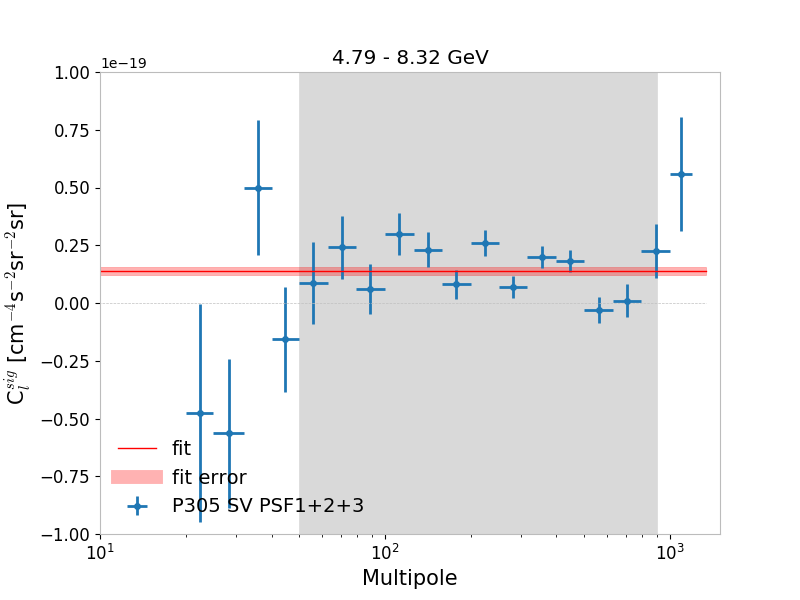}
\includegraphics[width=5.5cm]{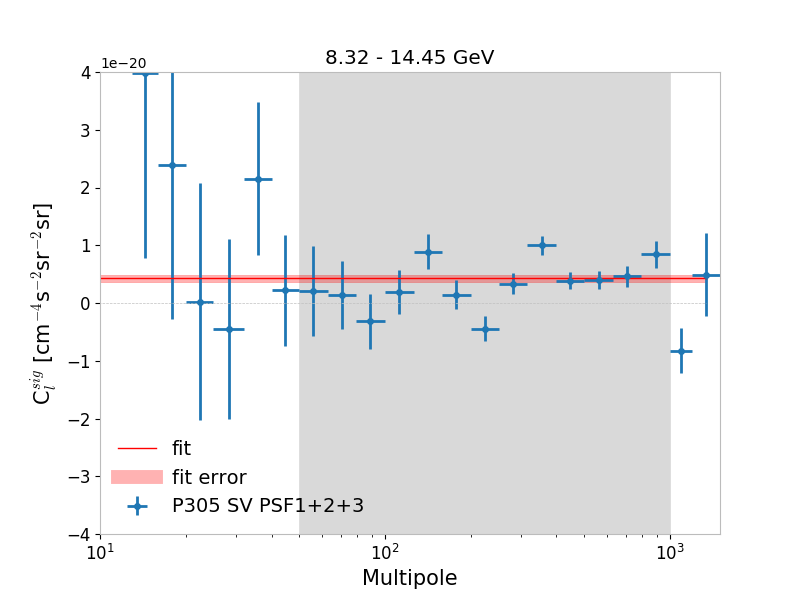}
\includegraphics[width=5.5cm]{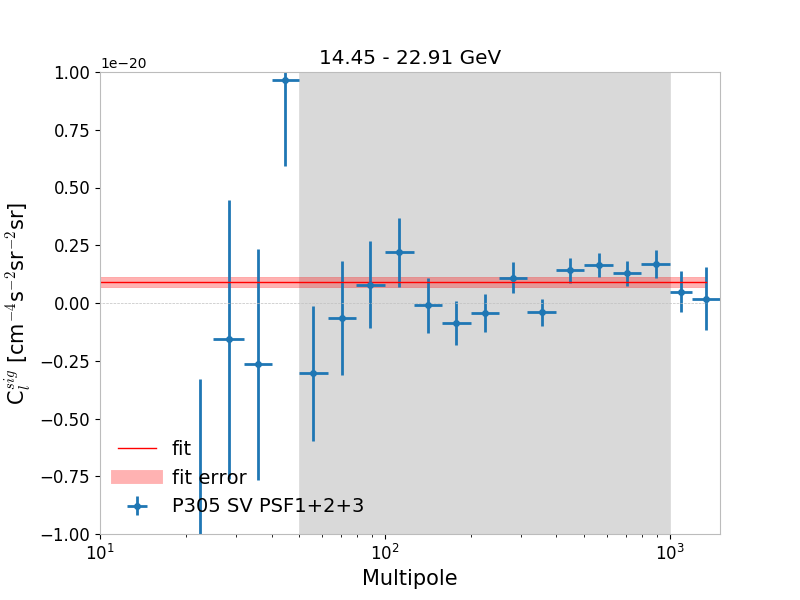}
\includegraphics[width=5.5cm]{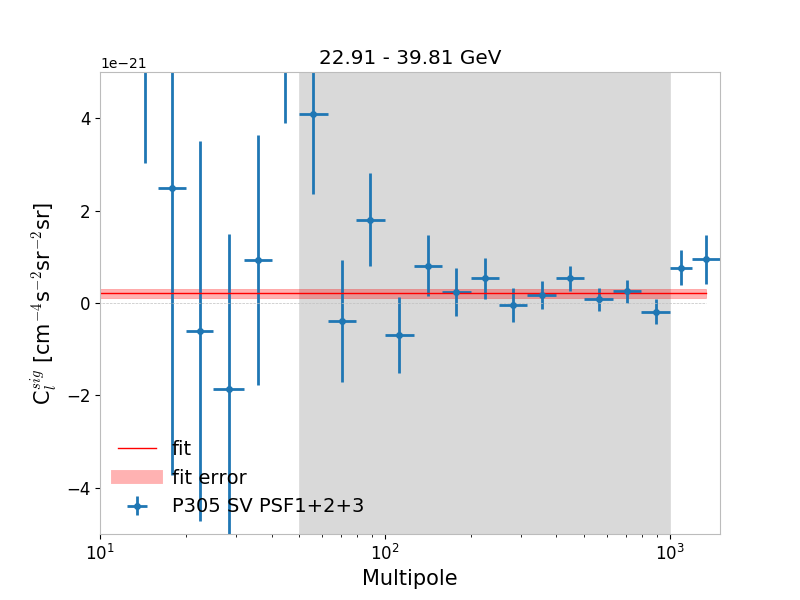}
\includegraphics[width=5.5cm]{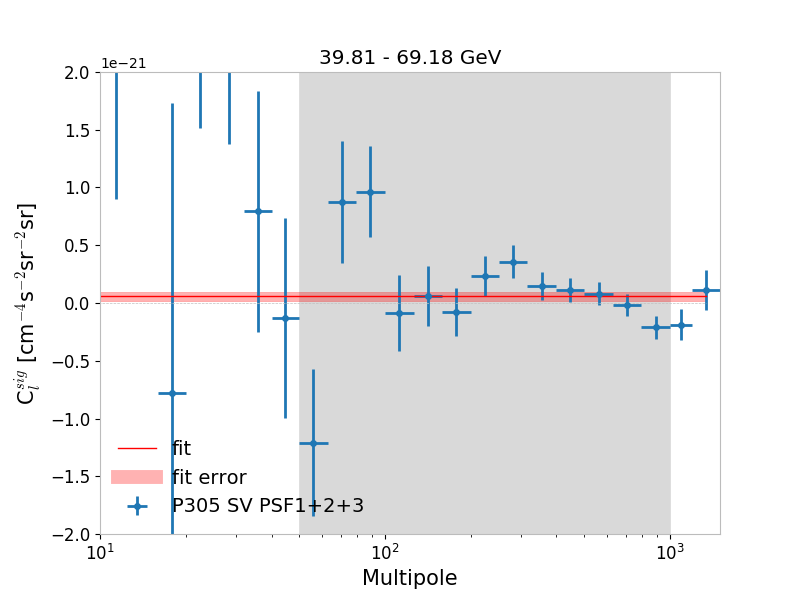}
\includegraphics[width=5.5cm]{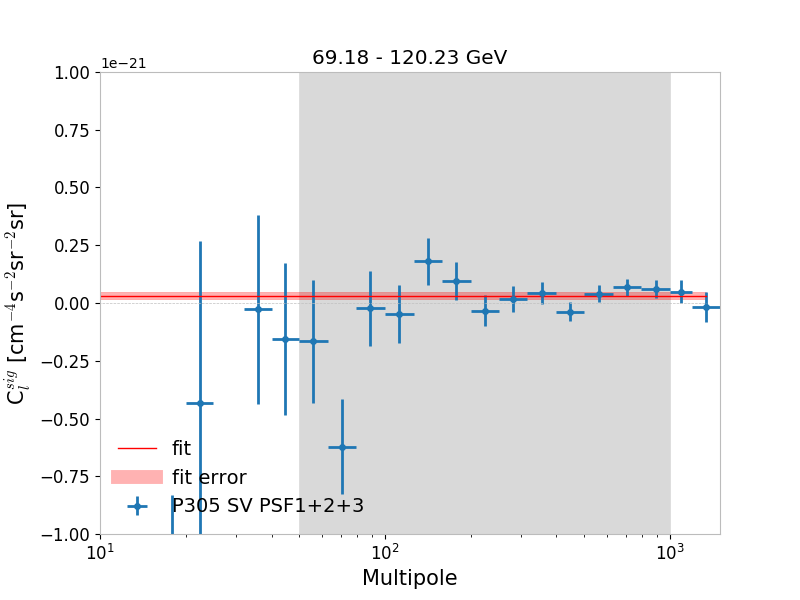}
\includegraphics[width=5.5cm]{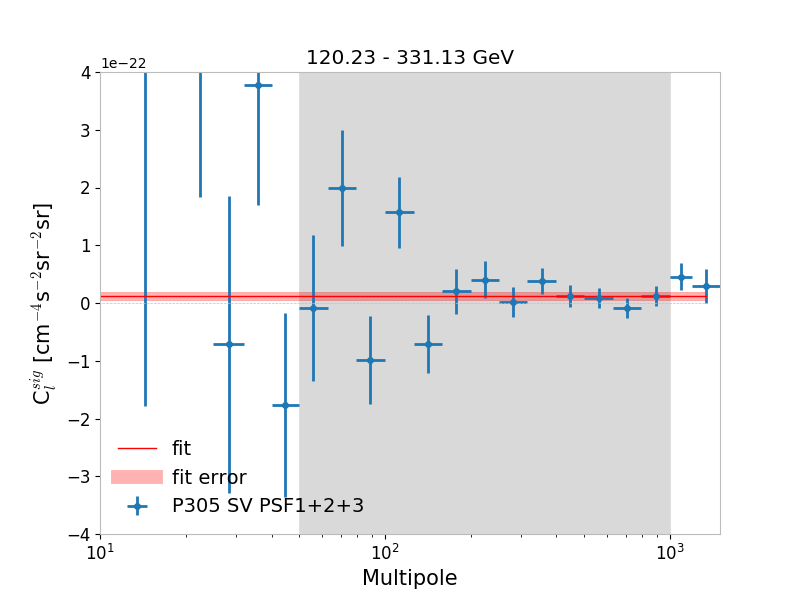}
\includegraphics[width=5.5cm]{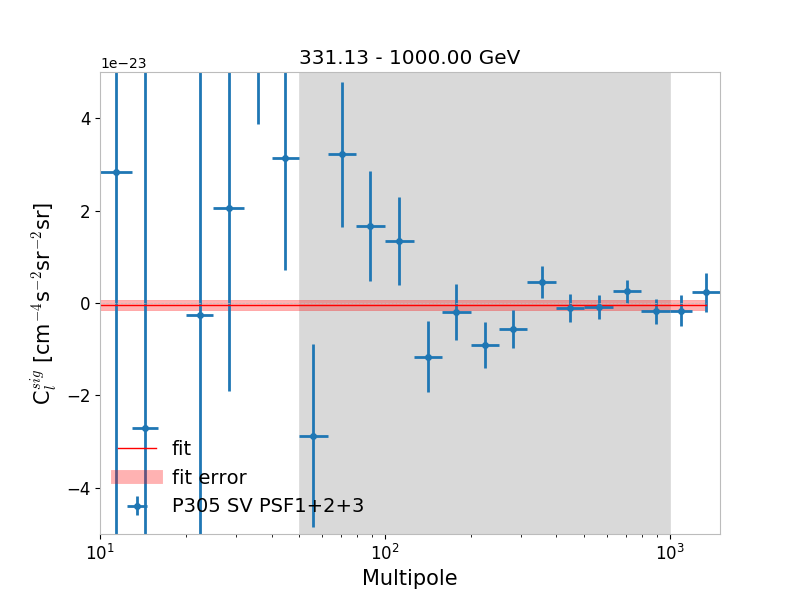}
\caption{Angular Power Spectra for all the energy bins. The shaded regions mark the ranges of multipoles considered in the fit of the APS to derive the anisotropy amplitudes $C_P$. The red lines show the C$_P$ values and their associated errors from the fit (represented by the red shaded band).}
\label{fig:AutoAPSfromecross}
\end{figure}
\end{addmargin}

\begin{addmargin}[-2em]{-2em}
\noindent
\begin{figure}[h]
\includegraphics[width=8cm]{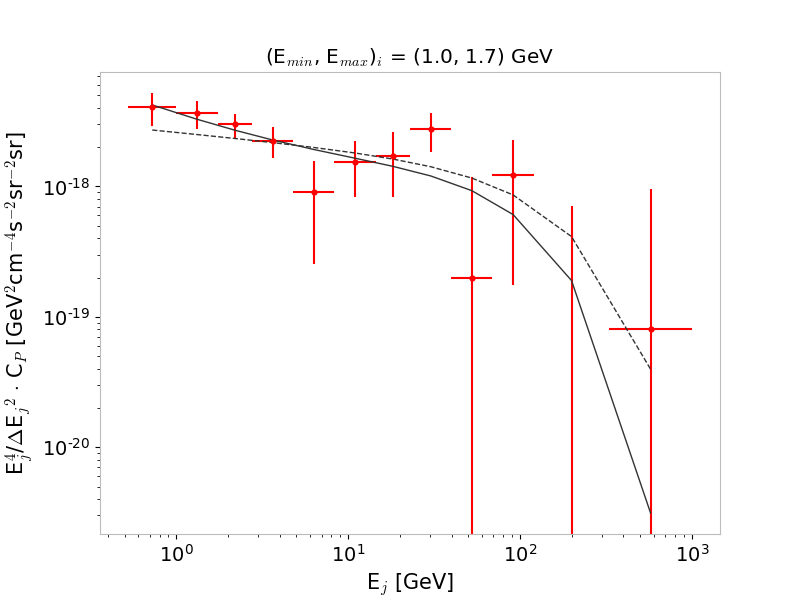}
\includegraphics[width=8cm]{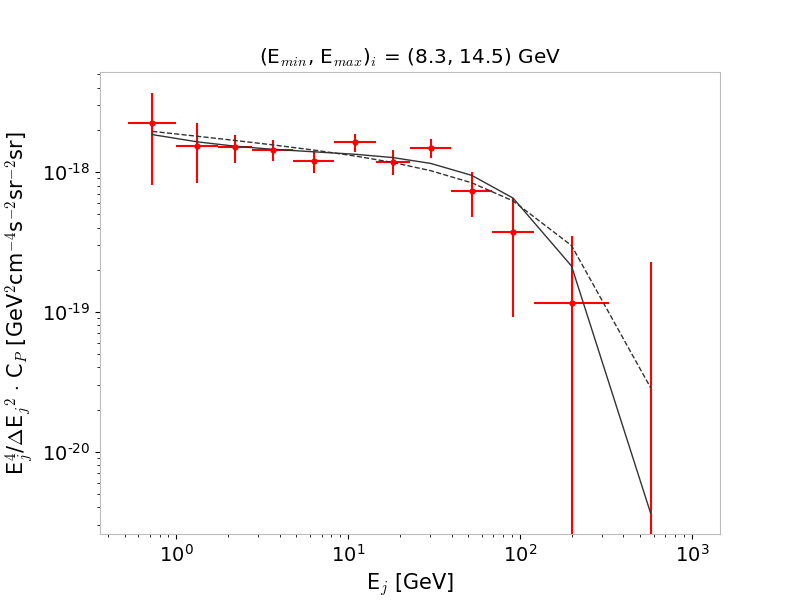}\\
\includegraphics[width=8cm]{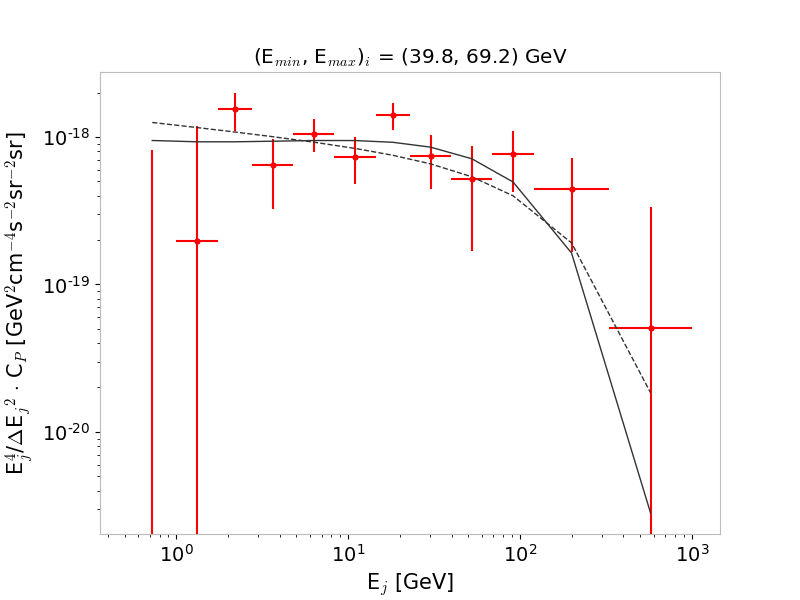}
\includegraphics[width=8cm]{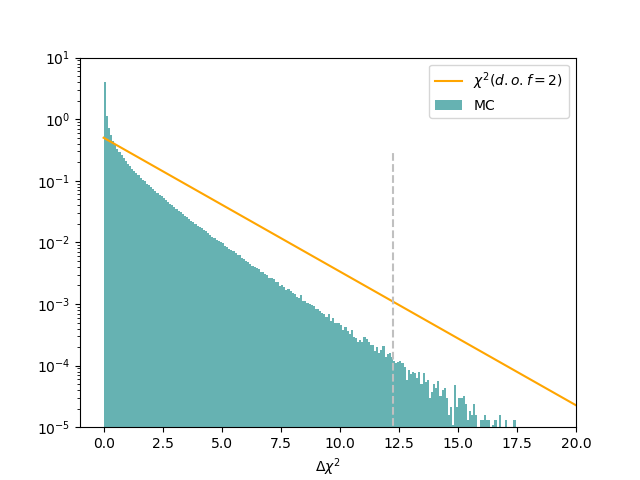}
\caption{Top and bottom-left: Cross-correlation anisotropy energy spectra for three energy bins: ($1.0-1.7$) GeV, ($8.3-14.5$) GeV and ($39.8-69.2$) GeV. The dashed line is the best-fit curve for the sPLE model, while the solid line is the best-fit curve for the dPLE model (see Sec. IV of the main paper for details about the fitting models). Bottom-right:  $\Delta\chi^2$ distribution obtained from the Monte Carlo procedure discussed in the text compared to a $\chi^2$ distribution with 2 degrees of freedom, which is the one predicted by the Wilks' theorem in the case of linear and nested models. Our model is nested but nonlinear, and the simulations show that the two additional degrees of freedom in the dPLE model result in a narrower distribution of $\Delta\chi^2$.}
\label{fig:cpij_fit}
\end{figure}
\end{addmargin}

\begin{figure}[h]
\centering
\includegraphics[width=8cm]{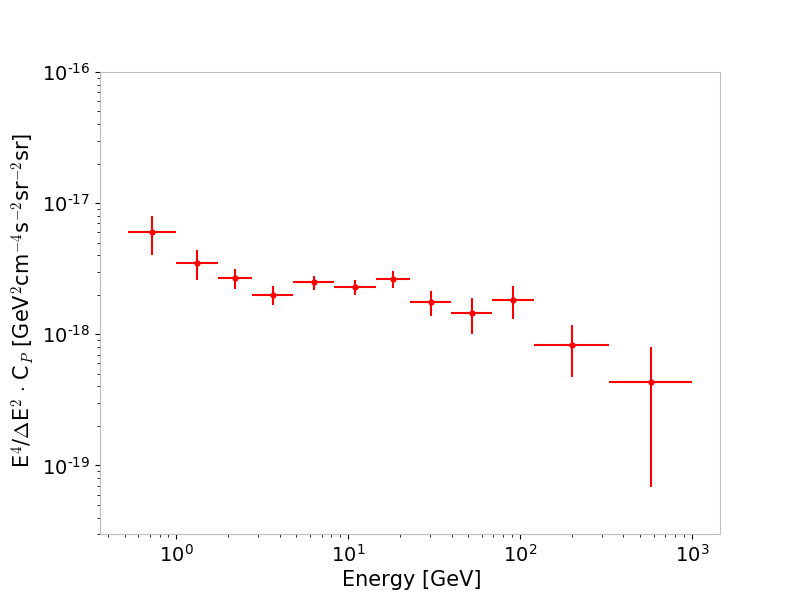}
\includegraphics[width=8cm]{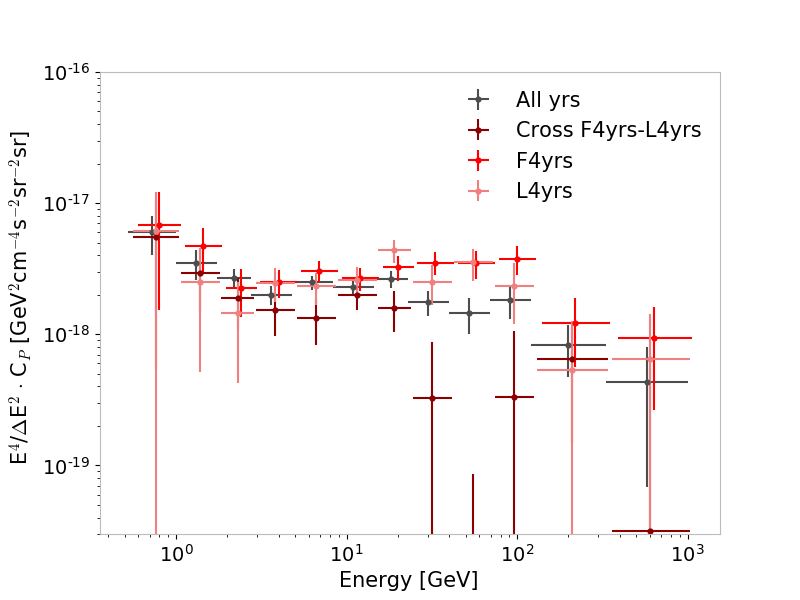}\\
\includegraphics[width=8cm]{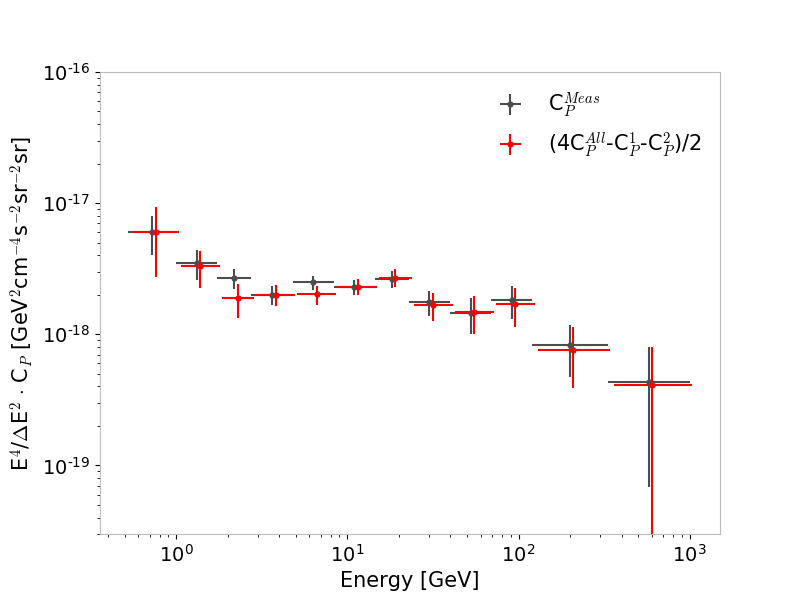}
\includegraphics[width=8cm]{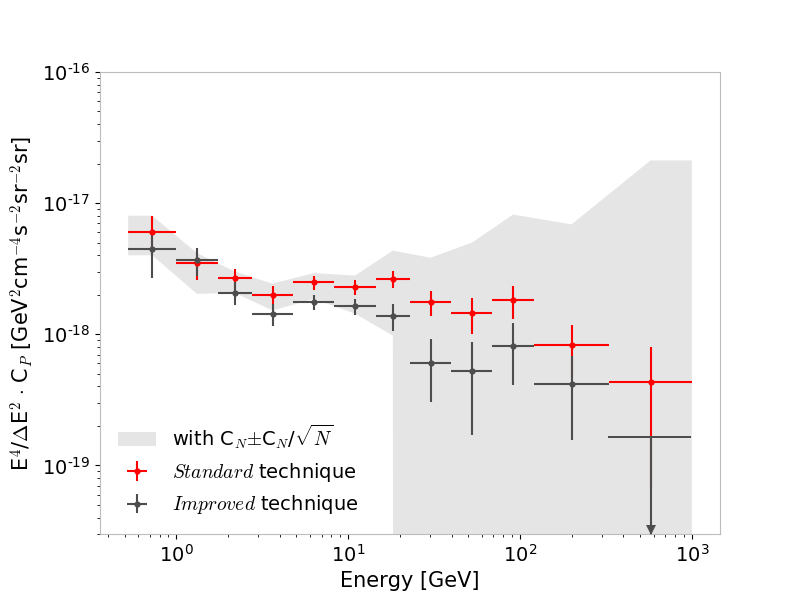}
\caption{Top-left: autocorrelation anisotropy energy spectrum resulting from the {\sl standard} autocorrelation analysis. Top-right: Together with the same spectrum shown in the top-left (gray points), we report the autocorrelation spectra for ``F4yrs" (red) and ``L4yrs" (pink) samples, and the cross-correlation spectrum between them (dark red). Bottom-left: Comparison between the autocorrelation anisotropy energy spectrum and $(C_{P, auto}^1 + C_{P, auto}^2 + 2C_{P, cross}^{1 2})/4$, where 1 and 2 indexes refer to the ``F4yrs" and ``L4yrs" samples. Bottom-right: Comparison between the anisotropy energy spectra obtained with the {\sl standard} autocorrelation analysis (gray) and the {\sl improved noiseless} correlation, obtained from the cross-correlation of the micro energy bins (red). The gray band is the statistical uncertainty associated with the white noise subtraction, given by $C_N/\sqrt{N.}$}
\label{fig:Cps_auto}
\end{figure}

\begin{figure}[h]
\centering
\includegraphics[width=11cm]{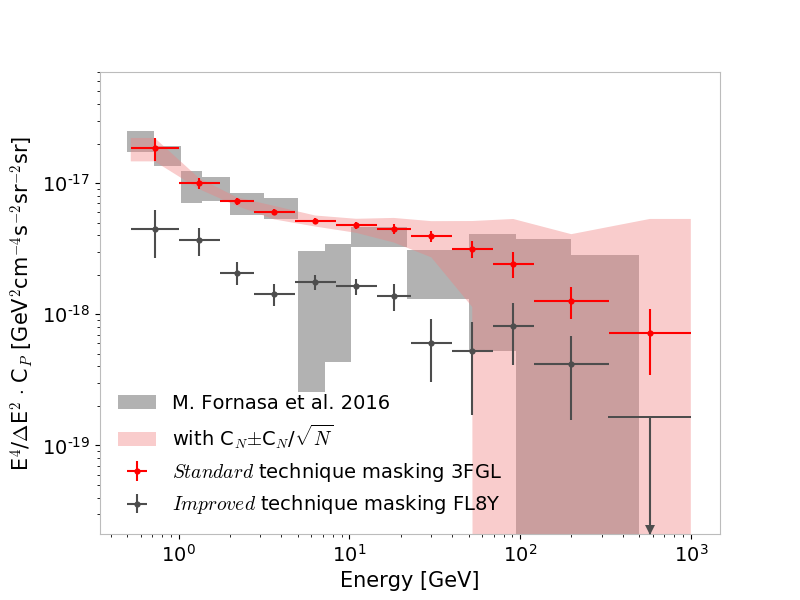}
\caption{Autocorrelation anisotropy energy spectrum obtained in this work (i.e., derived by using a point-source mask built on the FL8Y catalog and with the {\sl improved noiseless} technique) compared to the measurement performed by masking the 3FGL catalog and the {\sl standard} method of subtracting the Poisson noise. The latter is directly comparable to the previous analysis of Fornasa et al., shown by the gray squares, obtained with lower photon statistics.}
\label{fig:cp3FGL}
\end{figure}

\begin{figure}[h]
\centering
\includegraphics[width=8cm]{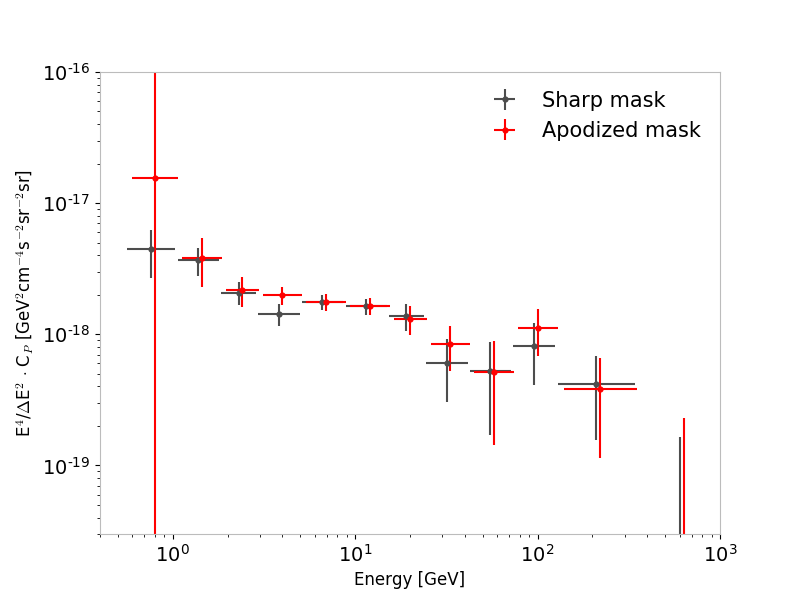}
\includegraphics[width=8cm]{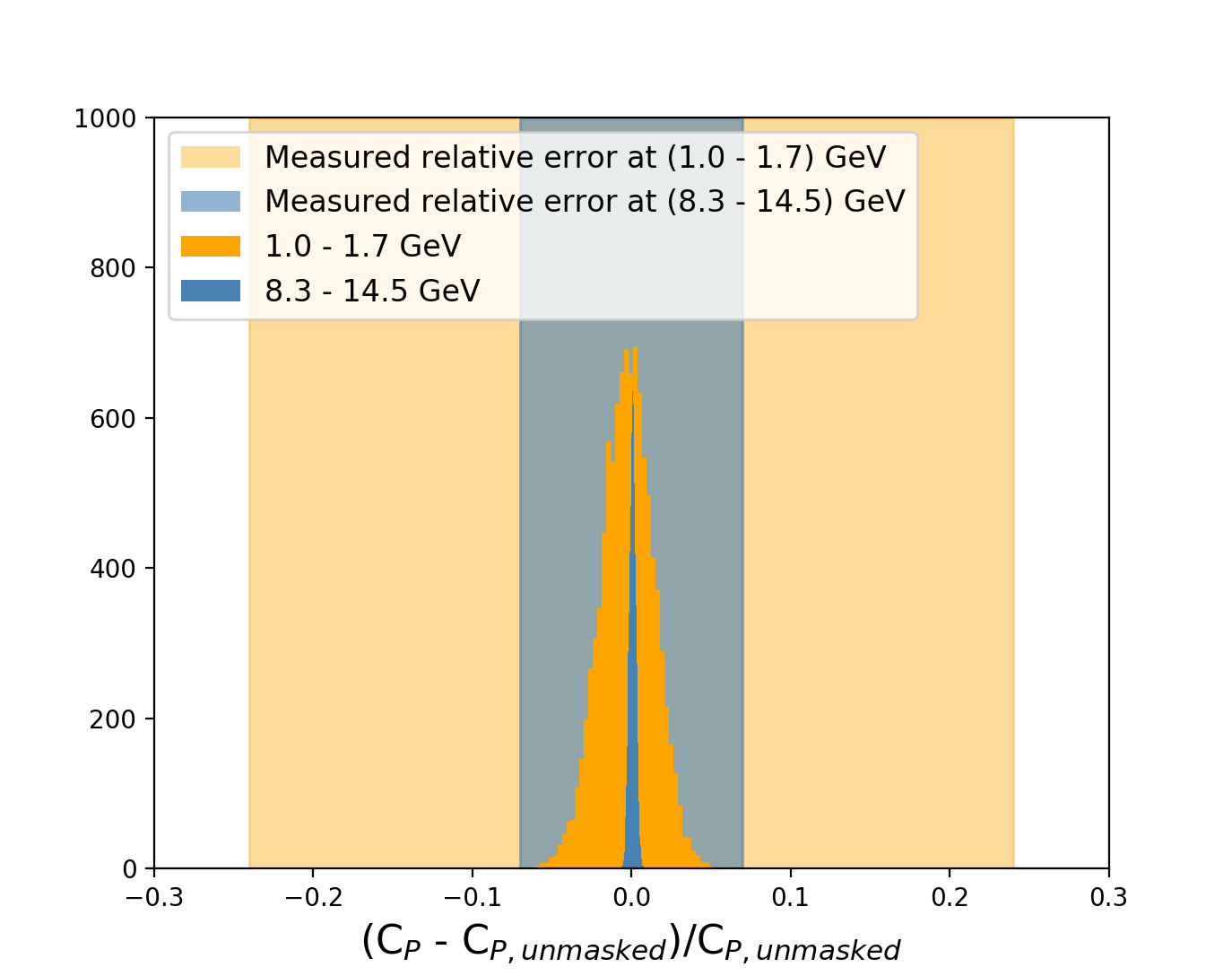}
\caption{Left: Comparison of the anisotropy energy spectra obtained with and without mask apodization. Right: Relative differences of the C$_P$ estimation, from {tt synfast}-simulated maps, with and without applying the mask (for bin ($1.-1.7$) GeV in orange and ($8.3-14.5$) GeV in blue). The bands illustrates the relative errors of the measured C$_P$ in the same energy bins.}
\label{fig:cpdiff_sim}
\end{figure}

\begin{sidewaystable}[h]
\centering
  \begin{tabular}{ c||c|c|c|c|c|c|c|c|c|c|c|c }
  \hline
 & \textbf{0.5-1.0} & \textbf{1.0-1.7 }& \textbf{1.7-2.8} & \textbf{2.8-4.8} & \textbf{4.8-8.3} & \textbf{8.3-14.5} & \textbf{14.5-22.9} & \textbf{22.9-39.8} & \textbf{39.8-62.2} & \textbf{62.2-120.2} & \textbf{120.2-331.1} & \textbf{331.1-1000.}\\
 &$[{\rm GeV}]$&$[{\rm GeV}]$&$[{\rm GeV}]$&$[{\rm GeV}]$&$[{\rm GeV}]$&$[{\rm GeV}]$&$[{\rm GeV}]$&$[{\rm GeV}]$&$[{\rm GeV}]$&$[{\rm GeV}]$&$[{\rm GeV}]$&$[{\rm GeV}]$\\
\hline
\hline

\textbf{0.5-1.0}& \cellcolor{lightgray}3.7E-18& & & & & & & & & & & \\
$[{\rm GeV}]$ &\cellcolor{lightgray}$\pm$1.5E-18& & & & & & & & & & & \\

\hline
\textbf{1.0-1.7}&1.6E-18& \cellcolor{lightgray}6.6E-19& & & & & & & & & & \\
$[{\rm GeV}]$ &$\pm$4.5E-19& \cellcolor{lightgray}$\pm$1.6E-19& & & & & & & & & & \\

\hline
\textbf{1.7-2.8}&9.2E-19& 2.7E-19&\cellcolor{lightgray}9.4E-20& & & & & & & & & \\
$[{\rm GeV}]$ &$\pm$2.3E-19&$\pm$5.6E-20&\cellcolor{lightgray}$\pm$1.8E-20& & & & & & & & & \\

\hline
\textbf{2.8-4.8}&3.7E-19&1.5E-19&8.2E-20&\cellcolor{lightgray}3.4E-20&& & & & & & & \\
$[{\rm GeV}]$ &$\pm$1.7E-19&$\pm$3.9E-20&$\pm$9.5E-21&\cellcolor{lightgray}$\pm$6.3E-21&& & & & & & & \\

\hline
\textbf{4.8-8.3}&2.1E-20&3.4E-20&2.8E-20&2.5E-20&\cellcolor{lightgray}1.4E-20& & & & & & & \\
$[{\rm GeV}]$ &$\pm$1.1E-19&$\pm$2.5E-20&$\pm$5.8E-21&$\pm$3.2E-21&\cellcolor{lightgray}$\pm$1.8E-21& & & & & & & \\

\hline
\textbf{8.3-14.5}&1.0E-19&3.3E-20&1.6E-20&1.1E-20&5.4E-21&\cellcolor{lightgray}4.3E-21& & & & & & \\
$[{\rm GeV}]$ &$\pm$6.6E-20&$\pm$1.5E-20&$\pm$3.6E-21&$\pm$1.9E-21&$\pm$9.6E-22&\cellcolor{lightgray}$\pm$6.1E-22& & & & & & \\

\hline
\textbf{14.5-22.9}&-5.2E-20&1.9E-20&4.8E-21&4.1E-21&2.5E-21&1.5E-21&\cellcolor{lightgray}9.0E-22&& & & & \\
$[{\rm GeV}]$ &$\pm$4.0E-20&$\pm$9.6E-21&$\pm$2.2E-21&$\pm$1.1E-21&$\pm$5.6E-22&$\pm$3.2E-22&\cellcolor{lightgray}$\pm$2.1E-22&& & & & \\

\hline
\textbf{22.9-39.8}&4.0E-20&2.2E-20&3.4E-21&4.4E-21&2.0E-21&1.4E-21&4.5E-22&\cellcolor{lightgray}2.1E-22& & & & \\
$[{\rm GeV}]$ &$\pm$3.0E-20&$\pm$7.1E-21&$\pm$1.6E-21&$\pm$8.4E-22&$\pm$4.0E-22&$\pm$2.3E-22&$\pm$1.3E-22&\cellcolor{lightgray}$\pm$1.0E-22& & & & \\

\hline
\textbf{39.8-62.2}&-1.1E-20&9.0E-22&3.5E-21&1.1E-21&1.0E-21&4.0E-22&3.8E-22&1.5E-22&\cellcolor{lightgray}5.9E-23&& & \\
$[{\rm GeV}]$ &$\pm$1.9E-20&$\pm$4.5E-21&$\pm$1.0E-21&$\pm$5.3E-22&$\pm$2.5E-22&$\pm$1.4E-22&$\pm$8.1E-23&$\pm$5.8E-23&\cellcolor{lightgray}$\pm$4.0E-23& & & \\

\hline
\textbf{62.2-120.2}&-1.1E-20&3.2E-21&1.3E-21&2.9E-22&3.7E-22&1.2E-22&7.8E-23&7.6E-23&5.0E-23&\cellcolor{lightgray}3.1E-23& & \\
$[{\rm GeV}]$ &$\pm$1.2E-20&$\pm$2.7E-21&$\pm$6.3E-22&$\pm$3.3E-22&$\pm$1.6E-22&$\pm$8.7E-23&$\pm$5.0E-23&$\pm$3.6E-23&$\pm$2.2E-23&\cellcolor{lightgray}$\pm$1.5E-23& & \\

\hline
\textbf{120.2-331.1}&1.1E-22&-3.2E-22&4.8E-22&7.2E-22&-4.1E-23&3.1E-23&-5.4E-23&2.2E-23&2.5E-23&9.7E-24&\cellcolor{lightgray}1.2E-23&\\
$[{\rm GeV}]$ &$\pm$8.0E-21&$\pm$1.9E-21&$\pm$4.4E-22&$\pm$2.2E-22&$\pm$1.1E-22&$\pm$6.2E-23&$\pm$3.5E-23&$\pm$2.5E-23&$\pm$1.6E-23&$\pm$9.7E-24&\cellcolor{lightgray}$\pm$7.3E-24&\\

\hline
\textbf{331.1-1000.}&1.2E-23&6.9E-23&1.5E-22&1.2E-22&-1.7E-23&-9.6E-25&-3.8E-25&4.5E-25&1.1E-24&-4.0E-24&-1.6E-24&\cellcolor{lightgray}-4.4E-25\\
$[{\rm GeV}]$ &$\pm$3.3E-21&$\pm$7.4E-22&$\pm$1.8E-22&$\pm$9.3E-23&$\pm$4.3E-23&$\pm$2.4E-23&$\pm$1.4E-23&$\pm$9.9E-24&$\pm$6.1E-24&$\pm$3.8E-24&$\pm$2.7E-24&\cellcolor{lightgray}$\pm$1.1E-24\\
 \hline
 \end{tabular}
\caption{Cross correlation amplitude C$_P^{ij}$ (in ${\rm cm}^{-4}{\rm s}^{-2}{\rm sr}^{-2}{\rm sr}$) for each pair of energy bins. The diagonal entries refer to the autocorrelation C$_P$.}
\label{tab:CpCross_val}
\end{sidewaystable}

\end{document}